\documentclass[envcountsect,envcountsame,oribibl]{llncs}
\usepackage{amssymb}
\usepackage[]{graphicx}
\usepackage[]{amsmath}
\usepackage{extarrows}
\usepackage[]{graphicx}
\usepackage{comment}
\usepackage[asymmetric,left=1in,top=1in,right=1in,bottom=0.5in]{geometry}

\title{Small Vertex Cover makes Petri Net Coverability and Boundedness
Easier}
\author{M. Praveen}
\institute{The Institute of Mathematical Sciences, Chennai, India}

\excludecomment{conf}
\includecomment{full}

\begin{document}
\pagestyle{plain}
\newcommand{\macname}[1]{\bf #1}
\newcommand{\macnat}{\mathbb{N}}

\newcommand{\macint}{\mathbb{Z}}

\newcommand{\macOh}{\mathcal{O}}

\newcommand{\macvec}[1]{\mathbf{#1}}

\newcommand{\macnet}{\mathcal{N}}
\newcommand{\macplaces}{P}
\newcommand{\macplacesu}{Q}
\newcommand{\mactranss}{T}
\newcommand{\macspplaces}{S}
\newcommand{\macidplaces}{I}
\newcommand{\macpre}{\mathrm{\mathit{Pre}}}
\newcommand{\macpost}{\mathrm{\mathit{Post}}}
\newcommand{\macarcw}{w}
\newcommand{\macmaxarcw}{W}
\newcommand{\macnumplaces}{m}
\newcommand{\macnumtrans}{n}
\newcommand{\macnumtranstype}{l}
\newcommand{\macnumtok}{e}
\newcommand{\macmaxcov}{R}
\newcommand{\macmaxinit}{U}

\newcommand{\EF}{\textbf{EF}}
\newcommand{\bdd}{< \omega}
\newcommand{\linf}{L}
\newcommand{\reach}{{\mathcal R}}
\newcommand{\nestDepth}{D}

\newcommand{\macplvar}{v}

\newcommand{\macmark}{M}
\newcommand{\macplaceo}{p}
\newcommand{\macplacet}{q}
\newcommand{\mactranso}{t}
\newcommand{\macStep}[1]{\xLongrightarrow{#1}}
\newcommand{\macstep}[2]{\xlongrightarrow[#2]{#1}}

\newcommand{\macfirseqo}{\sigma}
\newcommand{\macfso}{\sigma}
\newcommand{\macfirseqt}{\pi}
\newcommand{\macfst}{\pi}
\newcommand{\macfsh}{\eta}
\newcommand{\macfslidx}{r}

\newcommand{\macplaceidx}{i}
\newcommand{\macplaceidxt}{j}
\newcommand{\mactransidx}{i}
\newcommand{\mactypeidx}{j}
\newcommand{\macposition}{i}
\newcommand{\macposidx}{j}

\newcommand{\macvar}{\mathrm{\mathit{var}}}

\newcommand{\maclencov}{\mathrm{\mathit{lencov}}}
\newcommand{\maccovlen}{\ell}
\newcommand{\macslencov}{\mathrm{\mathit{slencov}}}
\newcommand{\macscovlen}{\ell_{1}}
\newcommand{\macidnidx}{j}

\newcommand{\macpupcv}{\macvec{b}}
\newcommand{\maclvs}{\macvec{L}}
\newcommand{\maclvm}{\macvec{B}}
\newcommand{\macef}{\mathrm{\mathit{\macvec{ef}}}}
\newcommand{\maclvidx}{i}
\newcommand{\macsolv}{\macvec{x}}
\newcommand{\macssolv}{\macvec{y}}

\newcommand{\maccc}[1]{\textsc{#1}}
\newcommand{\macpspace}{\maccc{Pspace}}
\newcommand{\macfpt}{\maccc{Fpt}}
\newcommand{\macwone}{\maccc{W[1]}}
\newcommand{\macwtwo}{\maccc{W[2]}}
\newcommand{\macxp}{\maccc{Xp}}
\newcommand{\macnp}{\maccc{Np}}
\newcommand{\macptime}{\maccc{Ptime}}
\newcommand{\macyes}{\maccc{Yes}}
\newcommand{\macparapsp}{\maccc{ParaPspace}}
\newcommand{\macexpsp}{\maccc{Expspace}}

\newcommand{\macpopterm}[1]{\textit{#1}}
\newcommand{\macstandout}[1]{{\bf #1}}

\newcommand{\Defref}[1]{Definition~\ref{#1}}
\newcommand{\defref}[1]{Def.~\ref{#1}}
\newcommand{\Thmref}[1]{Theorem~\ref{#1}}
\newcommand{\thmref}[1]{Theorem~\ref{#1}}
\newcommand{\Lemref}[1]{Lemma~\ref{#1}}
\newcommand{\lemref}[1]{Lemma~\ref{#1}}
\newcommand{\Propref}[1]{Proposition~\ref{#1}}
\newcommand{\propref}[1]{Prop.~\ref{#1}}
\newcommand{\Claimref}[1]{Claim~\ref{#1}}
\newcommand{\claimref}[1]{Claim~\ref{#1}}
\newcommand{\algoref}[1]{Algorithm~\ref{#1}}
\newcommand{\Algoref}[1]{Algorithm~\ref{#1}}
\newcommand{\Figref}[1]{Figure~\ref{#1}}
\newcommand{\figref}[1]{Fig.~\ref{#1}}
\newcommand{\Tabref}[1]{Table~\ref{#1}}
\newcommand{\tabref}[1]{Table~\ref{#1}}
\newcommand{\Secref}[1]{Section~\ref{#1}}
\newcommand{\secref}[1]{section~\ref{#1}}
\newcommand{\Subsecref}[1]{Sub-section~\ref{#1}}
\newcommand{\subsecref}[1]{Sub-section~\ref{#1}}
\newcommand{\Chref}[1]{Chapter~\ref{#1}}
\newcommand{\chref}[1]{Chap.~\ref{#1}}
\newcommand{\apndref}[1]{Appendix~\ref{#1}}
\newcommand{\Apndref}[1]{App.~\ref{#1}}

\newcommand{\macicon}{c}
\newcommand{\macicont}{d}
\newcommand{\maciconh}{h}

\newcommand{\macpoly}{\mathrm{\mathit{poly}}}

\newcommand{\macgraph}{G}
\newcommand{\macverto}{v}
\newcommand{\macvertext}{u}
\newcommand{\macedge}{e}
\newcommand{\macvertexs}{V}
\newcommand{\macedges}{E}
\newcommand{\maccols}{S}
\newcommand{\maccolor}{\ell}
\newcommand{\macvertcov}{VC}
\newcommand{\macvcnum}{k}

\newcommand{\macubs}{X}
\newcommand{\macubst}{Y}
\newcommand{\macnumpart}{\alpha}
\newcommand{\macpartidx}{\lambda}
\newcommand{\macpartidxt}{\mu}
\newcommand{\maceff}[1]{\Delta[#1]}
\newcommand{\macpump}[1]{\underline{#1}}
\newcommand{\macunbdd}{= \omega}
\newcommand{\maclinf}{L}
\newcommand{\macboundOnP}{c}
\newcommand{\macnumrep}{n}
\newcommand{\macpumlen}{\mathrm{\mathit{pumlen}}}
\newcommand{\macpumseqlen}{\ell_{2}}

\newcommand{\treenode}{\alpha}
\newcommand{\treenodes}{\Gamma}
\newcommand{\nodecontent}{content}
\newcommand{\boundfn}{f}
\newcommand{\wt}{\macmaxarcw}
\newcommand{\guessfn}{g}

\maketitle
\begin{abstract}
  The coverability and boundedness problems for Petri nets are known to
  be \macexpsp{}-complete. Given a Petri net, we associate a graph
  with it. With the vertex cover number $\macvcnum$ of this graph and the
  maximum arc weight $\macmaxarcw$ as parameters, we show that
  coverability and boundedness are in \macparapsp{}. This means that
  these problems can be solved in space
  $\macOh\left(\mathrm{\mathit{ef}}(\macvcnum,
  \macmaxarcw)\mathrm{\mathit{poly}}(n)\right)$, where
  $\mathrm{\mathit{ef}}(\macvcnum, \macmaxarcw)$ is some exponential
  function and $\mathrm{\mathit{poly}}(n)$ is some polynomial in the
  size of the input. We then extend the \macparapsp{} result to
  model checking a logic that can express some generalizations of
  coverability and boundedness.
\end{abstract}

\section{Introduction}
Petri nets, introduced by C. A. Petri \cite{Petri62}, are popularly
used for modelling concurrent infinite state systems. Using Petri
nets to verify various properties of concurrent systems is an ongoing
area of research, with abstract theoretical results like \cite{ABQ09}
and actually constructing tools for C programs like \cite{KMC02}.
Reachability, coverability and boundedness are some of the most
fundamental questions about Petri nets. All three of them are
\macexpsp{}-hard \cite{Lipton75}.  Coverability and boundedness are in
\macexpsp{} \cite{RCK78}.  Reachability is known to be decidable
\cite{Mayr81,Kos82} but no upper bound is known.

In this paper, we study the parameterized complexity of coverability
and boundedness problems. The parameters we consider are vertex cover
number $\macvcnum$ of the underlying graph of the given Petri net and
the maximum arc weight $\macmaxarcw$. We show that both problems can
be solved in space exponential in the parameters and polynomial in the
size of the input. Such algorithms are called \macparapsp{}
algorithms. Fundamental complexity theory of such parameterized
complexity classes have been studied \cite{FG03}, but parameterized
\macptime{} (popularly known as Fixed Parameter Tractable, \macfpt{})
is the most widely studied class. Usage of other parameterized classes
such as \macparapsp{} is rare in the literature.

As mentioned before, one of the uses of Petri nets is modelling
software. It is desirable to have better complexity bounds for certain
classes of Petri nets that may have some simple underlying structure
due to human designed systems that the nets model. For example, it is
known that well structured programs have small treewidth
\cite{Thorup98}. Unfortunately, the Petri net used by Lipton in the
reduction in \cite{Lipton75} (showing \macexpsp{}-hardness) has a
constant treewidth. Hence, we cannot hope to get better bounds for
coverability and boundedness with treewidth as parameter. Same is the
case with many other parameters like pathwidth, cycle rank, dagwidth
etc. Hence, we are forced to look for stronger parameters. In
\cite{PL09}, we studied the effect of a newly introduced parameter
called benefit depth. In this paper, we study the effect of using
vertex cover as parameter, using different techniques. The class of
Petri nets with bounded benefit depth is incomparable with the class
of Petri nets with bounded vertex cover.

Feedback vertex set of a graph is a set of vertices whose removal
leaves the graph without any cycles. The smallest feedback vertex set
of the Petri net used in the lower bound proof of \cite{Lipton75} is
large (as opposed to treewidth, pathwidth, cycle rank etc., which are
small). In the context of modelling software, smallest feedback vertex
set can be thought of as control points covering all loop structures.
In fact, the Petri net in the lower bound proof of \cite{Lipton75}
models a program that uses a large number of loops to manipulate
counters that can hold doubly exponential values. Removal of a
feedback vertex set leaves a Petri net without any cycles. It would be
interesting to explore the complexity of coverability and boundedness
problems with the size of the smallest feedback vertex set as
parameter. We have not been able to extend our results to the case of
feedback vertex set yet, but hope that these results will serve as a
theoretically interesting intermediate step.

In a tutorial article \cite{ESPCompl98}, Esparza argues that for most
interesting questions about Petri nets, the rule of thumb is that they
are all \macexpsp{}-hard. Despite this, the introduction of the same
article contains an excellent set of reasons for studying finer
complexity classification of such problems. We will not reproduce them
here but note some relevant points --- many experimental tools have
been built that solve \macexpsp{}-complete problems that can currently
handle small instances. Also, a knowledge of complexity of problems
helps in answering other questions. In such a scenario, having an
``extended dialog'' with the problem is beneficial, and parameterized
complexity is very good at doing this \cite{Downey03}.

\macpopterm{Related work.} In \cite{RY86}, Rosier and Yen study the
complexity of coverability and boundedness problems with respect to
different parameters of the input instance, such as number of places,
transitions, arc weight etc. In particular, they show that the space
required for boundedness is exponential in the number of unbounded
places and polynomial in the number of bounded places. If for a Petri
net, the smallest vertex cover is the set of all places, our results
coincide with those found in \cite{RY86}. Hence, our results refine
those of Rosier and Yen. In \cite{HAB97}, Habermehl
shows that the problem of model checking linear time $\mu$-calculus
formulas on Petri nets is \macpspace{}-complete in the size of the
formula and \macexpsp{}-complete in the size of the net. However, the
$\mu$-calculus considered in \cite{HAB97} cannot express coverability
and boundedness. In \cite{HCY92}, Yen extends the induction strategy
used by Rackoff in \cite{RCK78} to give \macexpsp{} upper bound for
deciding many other properties. Another work closely related to Yen's
above work is \cite{HA09}.

One-counter automata are closely related to Petri nets.
Precise complexity of reachability and many other problems of this
model have been recently obtained in \cite{HKOW09,GHOW10}. We have
adapted some of the techniques used in \cite{HKOW09,GHOW10}, in
particular the use of \cite[Lemma 42]{LLT04}.

The effect of treewidth and other parameters on the complexity of some
pebbling problems on digraphs have been considered in \cite[Section
5]{DFS99}. These problems relate to the reachability problem in a
class of Petri nets (called \emph{Elementary Net Systems}) with
semantics that are different from the ones used in this paper (see
\cite{RR98} for details of different Petri Net semantics).

\section{Preliminaries}
Let $\macint$ be the set of integers and $\macnat$ the set of natural
numbers. A Petri net is a 4-tuple $\macnet=(\macplaces,\mactranss,
\macpre,\macpost)$, where $\macplaces$ is  a set of places,
$\mactranss$ is a set of transitions and $\macpre$ and $\macpost$ are
the incidence functions: $\macpre:\macplaces\times \mactranss\to
\left[ 0\dots\macmaxarcw \right]$ (arcs going from places to
transitions) and $\macpost:\macplaces\times \mactranss\to \left[
0\dots\macmaxarcw\right]$ (arcs going from transitions to places),
where $\macmaxarcw\ge 1$. In diagrams, places will be represented by
circles and transitions by thick bars. Arcs are represented by
weighted directed edges between places and transitions.

A function $\macmark:\macplaces\to \macnat$ is called a
\macpopterm{marking}. A marking can be thought of as a configuration
of the Petri net, with every place $\macplaceo$ having
$\macmark(\macplaceo)$ tokens. Given a Petri net $\macnet$ with a
marking $\macmark$ and a transition $\mactranso$ such that for every
place $\macplaceo$, $\macmark(\macplaceo)\ge
\macpre(\macplaceo,\mactranso)$, the transition $\mactranso$ is said
to be \macpopterm{enabled} at $\macmark$ and can be
\macpopterm{fired}. After firing, the new marking $\macmark'$
(denoted as $\macmark\macStep{\mactranso}\macmark'$) is given
by
$\macmark'(\macplaceo)=\macmark(\macplaceo)-\macpre(\macplaceo,\mactranso)
+ \macpost(\macplaceo,\mactranso)$ for every place $\macplaceo$.  A
place $\macplaceo$ is an \macpopterm{input} (\macpopterm{output})
place of a transition $\mactranso$ if
$\macpre(\macplaceo,\mactranso)\ge 1$
($\macpost(\macplaceo,\mactranso)\ge 1$) respectively.  We can think
of firing a transition $\mactranso$ resulting in
$\macpre(\macplaceo,\mactranso)$ tokens being deducted from every
input place $\macplaceo$ and $\macpost(\macplaceo',\mactranso)$ tokens
being added to every output place $\macplaceo'$. A sequence of
transitions $\macfirseqo=\mactranso_{1}\mactranso_{2}\cdots
\mactranso_{\macfslidx}$ (called \macpopterm{firing sequence}) is said
to be enabled at a marking $\macmark$ if there are markings
$\macmark_{1},\dots,\macmark_{\macfslidx}$ such that
$\macmark\macStep{\mactranso_{1}}\macmark_{1}\macStep{\mactranso_{2}}
\cdots\macStep{\mactranso_{\macfslidx}} \macmark_{\macfslidx}$.
$\macmark, \macmark_{1},\dots,\macmark_{\macfslidx}$ are called
\macpopterm{intermediate markings}. The fact that firing $\macfirseqo$
at $\macmark$ results in $\macmark_{\macfslidx}$ is denoted by
$\macmark\macStep{\macfirseqo}\macmark_{\macfslidx}$.

\begin{figure}[!htp]
  \begin{center}
    \includegraphics{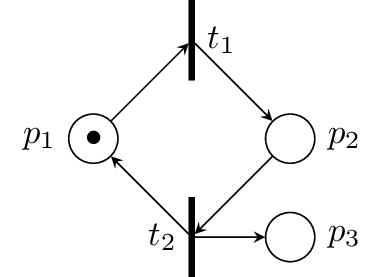}
  \end{center}
  \caption{An example of a Petri net}
  \label{fig:pnExample}
\end{figure}

We assume that a Petri net is presented as two matrices for $\macpre$
and $\macpost$. In the rest of this paper, we will assume that a Petri
net $\macnet$ has $\macnumplaces$ places, $\macnumtrans$ transitions
and that $\macmaxarcw$ is the maximum of the range of $\macpre$ and
$\macpost$. We define the size of the Petri net to be
$|\macnet|=2\macnumplaces\macnumtrans \log\macmaxarcw + \macnumplaces
\log|\macmark_{0}|$ bits, where $|M_{0}|$ is the maximum of the range
of the initial marking $\macmark_{0}$.
\begin{definition}[Coverability and Boundedness]
  \label{def:covBdd}
  Given a Petri net with an initial marking $\macmark_{0}$ and a
  target marking $\macmark_{cov}$, the \macpopterm{Coverability} problem
  is to determine if there is a firing sequence $\macfirseqo$ such that
  $\macmark_{0}\macStep{\macfirseqo}\macmark'$ and for every place
  $\macplaceo$, $\macmark'(\macplaceo)\ge \macmark_{cov}(\macplaceo)$
  (this is denoted as $\macmark'\ge \macmark_{cov}$). The boundedness
  problem is to determine if there is a number $\macicon\in \macnat$ such
  that for every firing sequence $\macfirseqo$ enabled at
  $\macmark_{0}$ with $\macmark_{0}\macStep{\macfirseqo}\macmark$,
  $\macmark(\macplaceo)\le \macicon$ for every place $\macplaceo$.
\end{definition}
In the Petri net shown in \figref{fig:pnExample}, the initial marking
$\macmark_{0}$ is given by $\macmark_{0}(\macplaceo_{1})=1$ and
$\macmark_{0}(\macplaceo_{2})=\macmark_{0}(\macplaceo_{3})=0$. If
$\macmark_{cov}$ is defined as
$\macmark_{cov}(\macplaceo_{1})=\macmark_{cov}(\macplaceo_{2})=1$ and
$\macmark_{cov}(\macplaceo_{3})=0$, then $\macmark_{cov}$ is not
coverable since $\macplaceo_{1}$ and $\macplaceo_{2}$ cannot have
tokens simultaneously. Since for any $\macicon\in \macnat$, the Petri
net in \figref{fig:pnExample} can reach a marking where
$\macplaceo_{3}$ has more than $\macicon$ tokens (by firing the
sequence $\mactranso_{1}\mactranso_{2}$ repeatedly), this Petri net is
not bounded. Lipton proved both coverability and boundedness problems
to be \macexpsp{}-hard \cite{Lipton75,ESPCompl98}.  Rackoff provided
\macexpsp{} upper bounds for both problems \cite{RCK78}. In the
definition of the coverability problem, if we replace $\macmark'\ge
\macmark_{cov}$ by $\macmark'=\macmark_{cov}$, we get the
\macpopterm{reachability} problem. Lipton's \macexpsp{} lower bound
applies to the reachability problem too, and this is the best known
lower bound.  Though the reachability problem is known to be decidable
\cite{Mayr81,Kos82}, no upper bound is known.
Many of the problems that are decidable for bounded Petri nets are
undecidable for unbounded Petri nets. Model checking some logics
extending the one defined in \secref{sec:logic} fall into this
category. Esparza and Nielsen survey such results in \cite{EspNiel94}.
Reachability, coverability and boundedness are few problems that
remain decidable for unbounded Petri nets.

\begin{conf}
  Proofs marked with (*) are skipped due to lack of space. A full
  version of this paper with the same title is available in arXiv,
  which contains all the proofs.
\end{conf}

\section{Vertex Cover for Petri Nets}
In this section, we introduce the notion of vertex cover for Petri
nets and intuitively explain how small vertex covers help in getting
better algorithms. We will also state and prove the key technical
lemma used in the next two sections.

For a normal graph $\macgraph=(\macvertexs,\macedges)$ with set of
vertices $\macvertexs$ and set of edges $\macedges$, a vertex cover
$\macvertcov\subseteq \macvertexs$ is a subset of vertices such that
every edge has at least one of its vertices in
$\macvertcov$. Given a Petri net $\macnet$, we associate with it an
undirected graph $\macgraph(\macnet)$ whose set of vertices is the set
of places $\macplaces$. Two vertices are connected by an edge if there
is a transition connecting the places corresponding to the two
vertices. To be more precise, if two vertices represent two places
$\macplaceo_{1}$ and $\macplaceo_{2}$, then there is an edge between
the vertices in $\macgraph(\macnet)$ iff in $\macnet$, there is some
transition $\mactranso$ such that
$\macpre(\macplaceo_{1},\mactranso)+\macpost(\macplaceo_{1},\mactranso)
\ge 1$ and
$\macpre(\macplaceo_{2},\mactranso)+\macpost(\macplaceo_{2},
\mactranso) \ge 1$. If a place $\macplaceo$ is both an input and
an output place of some transition, the vertex corresponding to
$\macplaceo$ has a self loop in $\macgraph(\macnet)$. Any vertex cover
of $\macgraph(\macnet)$ should include all vertices that have self
loops.

Suppose $\macvertcov$ is a vertex cover for some graph $\macgraph$. If
$\macverto_{1},\macverto_{2}\notin \macvertcov$ are two vertices not
in $\macvertcov$ that have the same set of neighbours (neighbours of a
vertex $\macverto$ are vertices that have an edge connecting
them to $\macverto$), $\macverto_{1}$ and $\macverto_{2}$ have
similar properties. This fact is used to obtain \macfpt{} algorithms for many
hard problems, e.g., see \cite{FLMRS08}. The same phenomenon leads to
\macparapsp{} algorithms for Petri net coverability and boundedness. In
the rest of this section, we will define the formalisms needed to
prove these results.

\begin{figure}[!htp]
  \begin{center}
    \includegraphics{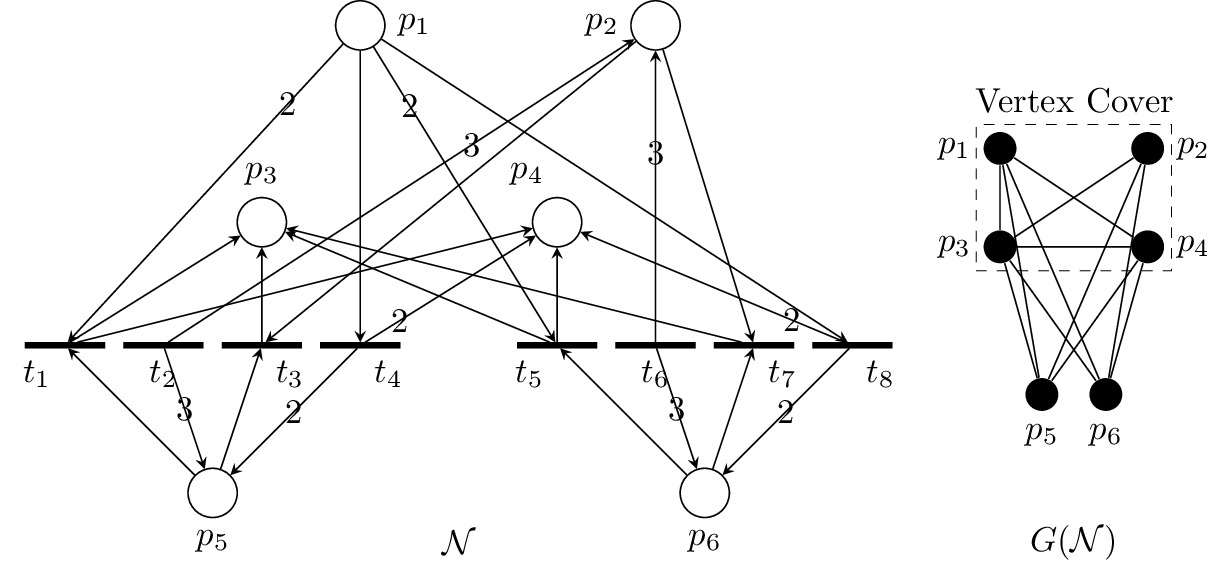}
  \end{center}
  \caption{A Petri net with vertex cover $\left\{
  \macplaceo_{1},\dots,\macplaceo_{4} \right\}$}
  \label{fig:vcExample}
\end{figure}

Let the places of a Petri net $\macnet$ be
$\macplaceo_{1},\macplaceo_{2},\dots,\macplaceo_{\macnumplaces}$.
Suppose there is a vertex cover $\macvertcov$ consisting of places
$\macplaceo_{1},\dots,\macplaceo_{\macvcnum}$. We say that two
transitions $\mactranso_{1}$ and $\mactranso_{2}$ are of the same type
if $\macpre(\macplaceo_{\macplaceidx},\mactranso_{1})=
\macpre(\macplaceo_{\macplaceidx},\mactranso_{2})$ and
$\macpost(\macplaceo_{\macplaceidx},\mactranso_{1})=\macpost(
\macplaceo_{\macplaceidx},\mactranso_{2})$ for all $\macplaceidx$
between $1$ and $\macvcnum$. In \figref{fig:vcExample}, transitions
$\mactranso_{1}$ and $\mactranso_{5}$ are of the same type.
Intuitively, two transitions of the same type behave similarly as far
as places in the vertex cover are concerned. Since there can be
$2\macvcnum$ arcs between a transition and places in $\macvertcov$ and
each arc can have weight between $0$ and $\macmaxarcw$, there can be
at most $(\macmaxarcw+1)^{2\macvcnum}$ different types of transitions.

Let $\macplaceo$ be a place not in the vertex cover $\macvertcov$.
Suppose there are $\macnumtranstype\le (\macmaxarcw+1)^{2\macvcnum}$
types of transitions. Place $\macplaceo$ can have one incoming arc
from or one outgoing arc to each transition of the net (it cannot have
both an incoming and an outgoing arc since in that case, $\macplaceo$
would have a self loop and would be in $\macvertcov$). If
$\macplaceo'$ is another place not in $\macvertcov$, then no
transition can have arcs to both $\macplaceo$ and $\macplaceo'$, since
otherwise, there would haven been an edge between $\macplaceo$ and
$\macplaceo'$ in $\macgraph(\macnet)$ and one of the places
$\macplaceo$ and $\macplaceo'$ would have been in $\macvertcov$.
Hence, places not in $\macvertcov$ cannot interact with each other
directly. Places not in $\macvertcov$ can only interact with places in
$\macvertcov$ through transitions and there are at most
$\macnumtranstype$ types of transitions. Suppose $\macplaceo$ and
$\macplaceo'$ have the following property: for every transition
$\mactranso$ that has an arc to/from $\macplaceo$ with weight
$\macarcw$, there is another transition $\mactranso'$ of the same type
as $\mactranso$ that has an arc to/from $\macplaceo'$ with weight
$\macarcw$.  Then, $\macplaceo$ and $\macplaceo'$ interact with
$\macvertcov$ in the same way in the following sense: whenever a
transition involving $\macplaceo$ fires, an ``equivalent'' transition
can be fired that involves $\macplaceo'$ instead of $\macplaceo$,
provided there are enough tokens in $\macplaceo'$. In
\figref{fig:vcExample}, places $\macplaceo_{5}$ and $\macplaceo_{6}$
satisfy the property stated above. Transition $\mactranso_{5}$ can be
fired instead of $\mactranso_{1}$, $\mactranso_{6}$ can be fired
instead of $\mactranso_{2}$ etc.
\begin{definition}
  \label{def:placeVariety}
  Suppose $\macnet$ is a Petri net with vertex cover $\macvertcov$ and
  $\macnumtranstype$ types of transitions. Let $\macplaceo\notin
  \macvertcov$ be a place not in the vertex cover. The variety
  $\macvar[\macplaceo]$ of $\macplaceo$ is defined as the
  function\footnote{The author acknowledges an anonymous IPEC referee
  for pointing out an error here in the submitted version.}
  $\macvar[\macplaceo]:\left\{ 1,\dots\macnumtranstype, \right\}\to
  2^{\left\{ -\macmaxarcw,\dots,\macmaxarcw \right\} \setminus
  \{0\}}$, where for every $\mactypeidx$ between $1$ and
  $\macnumtranstype$ and every $\macarcw \ne 0$ between $-\macmaxarcw$
  and $\macmaxarcw$, there is a transition $\mactranso_{\mactypeidx}$
  of type $\mactypeidx$ such that $\macarcw =
  -\macpre(\macplaceo,\mactranso_{\mactypeidx})+\macpost(\macplaceo,
  \mactranso_{\mactypeidx})$ iff $\macarcw \in \macvar[\macplaceo]$.
  We denote varieties of places by $\macplvar$, $\macplvar'$ etc.
\end{definition}
In the above definition, since $\macplaceo\notin \macvertcov$, at most
one among $\macpre(\macplaceo,\mactranso_{\mactypeidx})$ and
$\macpost(\macplaceo,\mactranso_{\mactypeidx})$ will be non-zero.

The fact that transitions can be exchanged between two places of the
same variety can be used to obtain better bounds on the length of
firing sequences. For example, suppose a firing sequence $\macfirseqo$
is fired in the Petri net of \figref{fig:vcExample}, with an initial
marking that has no tokens in $\macplaceo_{5}$ and $\macplaceo_{6}$. Let
$\macicon$ be the maximum number of tokens in any place in any
intermediate marking during the firing of $\macfirseqo$.
Since there are $6$ places and each intermediate marking has at most
$c$ tokens in every place, the number of possible distinct
intermediate markings is $(c+1)^{6}$. This is also an upper bound on the
length of $\macfirseqo$ (if two intermediate markings are
equal, then the subsequence between those two markings can be removed
without affecting the final marking reached). Now, suppose that in the final
marking reached, $\macplaceo_{5}$ and $\macplaceo_{6}$ do not have any
tokens and we replace all occurrences of
$\mactranso_{5},\mactranso_{6},\mactranso_{7}$ and $\mactranso_{8}$ in
$\macfirseqo$ by $\mactranso_{1},\mactranso_{2},\mactranso_{3}$ and
$\mactranso_{4}$ respectively. After this replacement, the final
marking reached will be same as the one reached after firing
$\macfirseqo$.
Number of tokens in $\macplaceo_{5}$ will be at most $2\macicon$ in
any intermediate marking and there will be no tokens at all in
$\macplaceo_{6}$. Variation in the number of tokens in
$\macplaceo_{1},\macplaceo_{2},\macplaceo_{3}$ and
$\macplaceo_{4}$ do not change (since as far as these places are
concerned, transitions $\mactranso_{5},\mactranso_{6},\mactranso_{7}$ and
$\mactranso_{8}$ behave in the same way as do
$\mactranso_{1},\mactranso_{2},\mactranso_{3}$ and
$\mactranso_{4}$ respectively). Hence, in any intermediate marking,
each of the places
$\macplaceo_{1},\macplaceo_{2},\macplaceo_{3}$ and
$\macplaceo_{4}$ will still have at most $\macicon$ tokens. When we
exchange the transitions as mentioned above, there might be some
intermediate markings that are same, so that we can get a shorter
firing sequence achieving the same effect as the original one. These
duplicate markings signify the ``redundancy'' that was present in the
original firing sequence $\macfirseqo$, but was not apparent to us due
to the distribution of tokens among places. After removing such
redundancies, the new upper bound on the length of the firing sequence
is $(2c+1).(c+1)^{4}$, which is asymptotically smaller than the previous bound
$(c+1)^{6}$. A careful observation of the effect of this phenomenon on
Rackoff's induction strategy in \cite{RCK78} leads us to the main
results of this paper.

\begin{definition}
  \label{def:firseqtransfer}
  Let $\macplaceo_{1}$ and $\macplaceo_{2}$ be two places of the same
  variety. Let $\macfirseqo$ be a firing sequence. A sequence of
  transitions
  $\macfirseqo'=\mactranso_{1}\dots\mactranso_{\macfslidx}$ is said to
  be a \macpopterm{sub-word} of $\macfirseqo$ if there are positions
  $\macposition_{1}<\cdots<\macposition_{\macfslidx}$ in $\macfirseqo$
  such that for each $\macposidx$ between $1$ and $\macfslidx$,
  $\macposition_{\macposidx}$\textsuperscript{th} transition of
  $\macfirseqo$ is $\mactranso_{\macposidx}$.
  Suppose $\macfirseqo'$ is a sub-word of $\macfirseqo$
  made up of transitions that have an arc to/from
  $\macplaceo_{1}$. \macstandout{Transferring $\macfirseqo'$ from
  $\macplaceo_{1}$ to $\macplaceo_{2}$} means replacing every
  transition $\mactranso$ of $\macfirseqo'$ (which has an arc to/from
  $\macplaceo_{1}$ with some weight $\macarcw$) with another
  transition $\mactranso'$ of the same type as $\mactranso$ which has
  an arc to/from $\macplaceo_{2}$ with weight $\macarcw$. The sub-word
  $\macfirseqo'$ is said to be \macstandout{safe for transfer} from
  $\macplaceo_{1}$ if for every prefix $\macfirseqo''$ of
  $\macfirseqo'$, the effect of $\macfirseqo''$ on $\macplaceo_{1}$
  (i.e., the change in the number of tokens in $\macplaceo_{1}$ as a
  result of firing all transitions in $\macfirseqo''$) is greater than
  or equal to $0$.
\end{definition}
Intuitively, if some sub-word $\macfirseqo'$ is safe for transfer from
$\macplaceo_{1}$, it never removes more tokens from $\macplaceo_{1}$
than it has already added to $\macplaceo_{1}$. So if we transfer
$\macfirseqo'$ from $\macplaceo_{1}$ to $\macplaceo_{2}$, the new
transitions will always add tokens to $\macplaceo_{2}$ before removing
them from $\macplaceo_{2}$, so there is no chance of number of tokens
in $\macplaceo_{2}$ becoming negative due to the transfer.  However,
the number of tokens in $\macplaceo_{1}$ may become negative due to
some old transitions remaining back in the ``untransferred'' portion
of the original firing sequence $\macfirseqo$. The following lemma
says that if some intermediate marking has very high number of tokens
in some place, then a suitable sub-word can be safely transfered
without affecting the final marking reached or introducing negative
number of tokens in any place, but reducing the maximum number of
tokens accumulated in any intermediate marking. The proof is a simple
consequence of \cite[Lemma 42]{LLT04}, which is about one-counter
automata.
\begin{full}
An one-counter automaton is an automaton with a counter that can store
natural numbers. Apart from changing its state, the automaton can
increment the counter, test it for zero and decrement it when not
zero. It is proven in \cite[Lemma 42]{LLT04} that if a one-counter
automaton can reach from one of its configuration to another, it can
do so without increasing the intermediate values of the counter by
large numbers.  A full proof of the following lemma is included in the
Appendix for easy reference.
\end{full}
\begin{conf}
  A full proof of the following lemma can be found in the Appendix of
  the full version.
\end{conf}

\begin{lemma}[Truncation lemma, \cite{LLT04}]
  \label{lem:truncationLemma}
  Let $\macplaceo_{1}$ and $\macplaceo_{2}$ be places of the same
  variety. Let $\macnumtok\in \macnat$ be any number and $\macfirseqo$
  be a firing sequence. Suppose during the firing of $\macfirseqo$,
  there are intermediate markings $\macmark_{1}$ and $\macmark_{3}$
  such that $\macmark_{1}(\macplaceo_{1})= \macnumtok$  and
  $\macmark_{3}(\macplaceo_{1})\le \macnumtok$. Suppose $\macmark_{2}$
  is an intermediate marking between $\macmark_{1}$ and $\macmark_{3}$
  such that $\macmark_{2}(\macplaceo_{1})\ge
  \macnumtok+\macmaxarcw^{2}+\macmaxarcw^{3}$ is the maximum number of
  tokens in $\macplaceo_{1}$ at any intermediate marking between
  $\macmark_{1}$ and $\macmark_{3}$. Then, there is a sub-word
  $\macfirseqo'$ of $\macfirseqo$ that is safe for transfer from
  $\macplaceo_{1}$ to $\macplaceo_{2}$ such that
  \begin{enumerate}
    \item The total effect of $\macfirseqo'$ on $\macplaceo_{1}$ is
      $0$.
    \item After transferring $\macfirseqo'$ to $\macplaceo_{2}$, the
      number of tokens in $\macplaceo_{1}$ at $\macmark_{2}$ is
      strictly less than the number of tokens in $\macplaceo_{1}$ at
      $\macmark_{2}$ before the transfer.
    \item No intermediate
      marking will have negative number of tokens in $\macplaceo_{1}$
      after the transfer.
  \end{enumerate}
\end{lemma}

There can be at most $(2^{2\macmaxarcw})^{\macnumtranstype}\le
2^{2\macmaxarcw (\macmaxarcw+1)^{2\macvcnum}}$ varieties of places that
are not in the vertex cover $\macvertcov$, if the number of places in
the vertex cover is $\macvcnum$. For each variety $\macplvar$, we
designate one of the places having $\macplvar$ as its variety as special,
and use $\macplaceo_{\macplvar}$ to denote it. We will call
$\macspplaces = \macvertcov\cup \left\{ \macplaceo_{\macplvar}\mid
\macplvar\text{ is the variety of a place not in }\macvertcov
\right\}$   the set of special places. We will
denote the set $\macplaces\setminus \macspplaces$ using
$\macidplaces$ and call the places in $\macidplaces$ independent
places. We will use $\macvcnum'$ for the cardinality of $\macspplaces$
and note that $\macvcnum'\le
\macvcnum+2^{2\macmaxarcw (\macmaxarcw+1)^{2\macvcnum}}$. If
$\macvcnum$ and $\macmaxarcw$ are parameters, then $\macvcnum'$ is a
function of the parameters only. Hence, in the rest of the paper, we
will treat $\macvcnum'$ as the parameter.

\section{\macparapsp{} algorithm for the Coverability problem}

In this section, we will show that for a Petri net $\macnet$ with
a vertex cover of size $\macvcnum$ and maximum arc weight $\macmaxarcw$,
the coverability problem can be solved in space
$\macOh(\mathrm{\mathit{ef}}(\macvcnum,\macmaxarcw)\mathrm{\mathit{poly}}(|\macnet|+\log|\macmark_{cov}|))$.
Here, $\mathrm{\mathit{ef}}$ is some computable function exponential
in $\macvcnum$ and $\macmaxarcw$ while
$\mathrm{\mathit{poly}}(|\macnet|+\log|\macmark_{cov}|)$ is some
polynomial in the size of the net and the marking to be covered. We
will need the following definition, which is Definition 3.1 from
\cite{RCK78} adapted to our notation.

\begin{definition}
  \label{def:QCovSeq}
  Let $\macplacesu\subseteq \macplaces$ be some subset of places such
  that $\macidplaces\subseteq \macplacesu$. For a transition
  $\mactranso$ and functions $\macmark,\macmark':\macplaces\to
  \macint$, we write $\macmark\macstep{\mactranso}{\macplacesu}\macmark'$
  if
  $\macmark'(\macplaceo)=\macmark(\macplaceo)-\macpre(\macplaceo,\mactranso)
  + \macpost(\macplaceo,\mactranso)$ for all $\macplaceo\in
  \macplaces$ and $\macmark(\macplacet),\macmark'(\macplacet)\ge 0$
  for all $\macplacet\in \macplacesu$. Let $\macmark_{cov}$ be
  some marking to be covered. For a function
  $\macmark_{0}:\macplaces\to \macint$, a firing sequence
  $\macfirseqo=\mactranso_{1}\mactranso_{2}\cdots\mactranso_{\macfslidx}$
  is said to be \macstandout{$\macplacesu$-covering
  from $\macmark_{0}$} if there are intermediate functions
  $\macmark_{1},\macmark_{2},\dots,\macmark_{\macfslidx}$ such that
  $\macmark_{0}\macstep{\mactranso_{1}}{{\macplacesu}}\macmark_{1}
  \macstep{\mactranso_{2}}{\macplacesu} \cdots
  \macstep{\mactranso_{\macfslidx}}{\macplacesu}\macmark_{\macfslidx}$
  and $\macmark_{\macfslidx}(\macplacet)\ge
  \macmark_{cov}(\macplacet)$ for all $\macplacet\in \macplacesu$. The
  firing sequence $\macfirseqo$ is further said to be
  $\macplacesu,\macnumtok$-covering if for all $\mactransidx$ between
  $0$ and $\macfslidx-1$, the functions $\macmark_{\mactransidx}$
  above satisfy $\macmark_{\mactransidx}(\macplacet)\le \macnumtok$
  for all $\macplacet\in \macplacesu$. For a function
  $\macmark:\macplaces\to \macint$, let
  $\maclencov(\macplacesu,\macmark, \macmark_{cov})$ be the length of the shortest
  firing sequence that is $\macplacesu$-covering from $\macmark$.
  Define $\maclencov(\macplacesu,\macmark, \macmark_{cov})$ to be $0$ if there is no
  such sequence. Define $\maccovlen(\macplaceidx)=\max\left\{
  \maclencov(\macplacesu,\macmark, \macmark_{cov})\mid \macidplaces\subseteq
  \macplacesu\subseteq \macplaces, |\macplacesu\setminus\macidplaces|=
  \macplaceidx,\macmark:\macplaces\to \macint \right\}$.
\end{definition}
Intuitively, a $\macplacesu$-covering sequence does not care about
places that are not in $\macplacesu$, even if some intermediate
markings have ``negative number of tokens''. The number
$\maccovlen(\macplaceidx)$ is an upper bound on the length of covering
sequences that only care about independent places and $\macplaceidx$
special places. Obviously, we are only interested in
$\maccovlen(\macvcnum')$, but other values help in obtaining it. With
slight abuse of terminology, we will call functions
$\macmark:\macplaces\to \macint$ also as markings. It will be clear
from context what is meant.

Let $\macmaxcov$ be the maximum of the range of $\macmark_{cov}$, the
marking to be covered. We will denote
$\macmaxcov+\macmaxarcw+\macmaxarcw^{2}+\macmaxarcw^{3}$ by
$\macmaxcov'$. Recall that $\macnumplaces$ is the number of places in
the given Petri net.  The following lemmas give an upper bound on
$\maccovlen(\macvcnum')$.
\begin{lemma}
  \label{lem:covLenBaseCase}
  $\maccovlen(0)\le \macnumplaces\macmaxcov$.
\end{lemma}
\begin{proof}
  $\maccovlen(0)$ is the length of the shortest
  $\macidplaces$-covering sequence. Recall that all places in
  $\macidplaces$ are independent of each other, so if a transition has
  an arc to one of the places in $\macidplaces$, it does not have arcs
  to any other place in $\macidplaces$. Since an
  $\macidplaces$-covering sequence does not care about places in
  $\macspplaces$, it only has to worry about adding tokens to places
  in $\macidplaces$. If a transition adds a token to some place
  $\macplaceo$ in $\macidplaces$, it does not remove tokens from any
  other place in $\macidplaces$. Hence, this transition can be
  repeated $\macmaxcov$ times to add at least $\macmaxcov$ tokens to
  the place $\macplaceo$, which is all that is needed for
  $\macplaceo$. Arguing similarly for other places in $\macidplaces$,
  a total of $\macnumplaces\macmaxcov$ transitions are enough to add
  all required tokens to all places in $\macidplaces$, since there are
  less than $\macnumplaces$ places in $\macidplaces$.\qed
\end{proof}
\begin{lemma}
  \label{lem:covLenIndStep}
  $\maccovlen(\macplaceidx+1)\le
  \macmaxcov'^{m}(\macmaxarcw\maccovlen(\macplaceidx) +
  \macmaxcov)^{\macplaceidx+1} + \maccovlen(\macplaceidx)$.
\end{lemma}
\begin{full}
\begin{proof}
  Suppose $\macidplaces\subseteq \macplacesu\subseteq \macplaces$ and
  $|\macplacesu\setminus \macidplaces|=\macplaceidx+1$. Suppose there
  is a sequence $\macfirseqo$ that is $\macplacesu$-covering from some
  $\macmark_{0}$. Let $\macplaceo$ be any place in $\macidplaces$ of
  some variety $\macplvar$. Let $\macmark$ be the first intermediate
  marking such that $\macmark(\macplaceo)\ge
  \macmark_{cov}(\macplaceo)$. We have $\macmark(\macplaceo)\le
  \macmaxcov+\macmaxarcw$. We distinguish two cases:
  \begin{enumerate}
    \item For all intermediate markings $\macmark'$ after $\macmark$,
      $\macmark'(\macplaceo)\ge \macmark(\macplaceo)$. This means the
      number of tokens in $\macplaceo$ never goes below
      $\macmark(\macplaceo)$ after the marking $\macmark$. Let
      $\macfirseqo'$ be the sub-word of $\macfirseqo$ that consists
      of all transition occurrences after $\macmark$ that has an arc
      to/from $\macplaceo$. The sub-word $\macfirseqo'$ is safe for
      transfer from $\macplaceo$ to $\macplaceo_{\macplvar}$. We
      transfer $\macfirseqo'$ from $\macplaceo$ to
      $\macplaceo_{\macplvar}$ and note that in the final marking
      reached after the transfer, $\macplaceo$ still has
      $\macmark(\macplaceo)$ tokens, which is enough to cover
      $\macmark_{cov}$.
    \item Let $\macmark'$ be the last intermediate marking such that
      $\macmark'(\macplaceo)<\macmark(\macplaceo)$. We invoke the
      truncation lemma by setting
      $\macnumtok=\macmark(\macplaceo)\le \macmaxcov+\macmaxarcw$,
      $\macmark_{1}=\macmark$ and $\macmark_{3}=\macmark'$. We can
      then transfer the sub-word $\macfirseqo'$ identified by the
      truncation lemma to $\macplaceo_{\macplvar}$ to reduce the
      number of tokens in $\macplaceo$ in some intermediate
      markings between $\macmark$ and $\macmark'$. We repeat this
      process until there are no more than $\macmaxcov'$ tokens in
      $\macplaceo$ in any intermediate marking between $\macmark$ and
      $\macmark'$. Let $\macmark''$ be the first intermediate marking
      after $\macmark'$ such that $\macmark''(\macplaceo) \ge
      \macmark_{cov} (\macplaceo)$. Again, $\macmark'' ( \macplaceo)
      \le \macmaxcov + \macmaxarcw$. If no intermediate marking
      $\macmark_{3}''$ after $\macmark''$ has $\macmark_{3}''
      ( \macplaceo) < \macmark'' ( \macplaceo)$, we can transfer all
      transitions with an arc to/from $\macplaceo$ occurring after
      $\macmark''$ to $\macplaceo_{\macplvar}$. Otherwise, we can
      invoke truncation lemma again to ensure that
      $\macplaceo$ has at most $\macmaxcov'$ tokens in any
      intermediate marking after $\macmark''$.
  \end{enumerate}
  Repeating the above case analysis for every independent place
  $\macplaceo\in \macidplaces$, we get a firing sequence $\macfirseqt$
  that is $\macplacesu$-covering from $\macmark_{0}$ such that in all
  intermediate markings, every independent place $\macplaceo$ has at
  most $\macmaxcov'$ tokens. If this sequence happens to be
  $\macplacesu,(\macmaxarcw\maccovlen(\macplaceidx)+\macmaxcov)$-bounded,
  then
  $\macmaxcov'^{m}(\macmaxarcw\maccovlen(\macplaceidx)+\macmaxcov)^{\macplaceidx+1}$
  is an upper bound on its length (since all independent places have
  at most $\macmaxcov'$ tokens and the $\macplaceidx+1$ places in
  $\macplacesu\setminus \macidplaces$ have at most
  $(\macmaxarcw\maccovlen(\macplaceidx)+\macmaxcov)$ tokens in all
  intermediate markings) and we are done.
  
  Otherwise, suppose there is some place $\macplacet\in
  \macplacesu\setminus \macidplaces$ and some intermediate marking
  $\macmark$ such that $\macmark(\macplacet)\ge
  \macmaxarcw\maccovlen(\macplaceidx)+\macmaxcov$. Let $\macmark$ be
  the first such marking and call the prefix of $\macfirseqt$ up to
  $\macmark$ as $\macfirseqt_{1}$ and the rest of $\macfirseqt$ as
  $\macfirseqt_{2}$. The length of $\macfirseqt_{1}$ is at most
  $\macmaxcov'^{m}(\macmaxarcw\maccovlen(\macplaceidx)+\macmaxcov)^{\macplaceidx+1}$.
  The sequence $\macfirseqt_{2}$ is a $(\macplacesu\setminus\left\{ \macplacet
  \right\})$-covering sequence from $\macmark$. By definition, there is
  such a sequence $\macfirseqt_{2}'$ of length at most
  $\maccovlen(\macplaceidx)$.  The sequence $\macfirseqt_{1}\macfirseqt_{2}'$
  is a $(\macplacesu\setminus\left\{ \macplacet \right\})$-covering
  sequence from $\macmark_{0}$. Since $\macmark(\macplacet)\ge
  \macmaxarcw\maccovlen(\macplaceidx)+\macmaxcov$ and
  $\macfirseqt_{2}'$ removes at most
  $\macmaxarcw\maccovlen(\macplaceidx)$ tokens from $\macplacet$,
  $\macfirseqt_{1}\macfirseqt_{2}'$ is in fact a
  $\macplacesu$-covering sequence from $\macmark_{0}$. Its length is
  bounded by
  $\macmaxcov'^{m}(\macmaxarcw\maccovlen(\macplaceidx)+\macmaxcov)^{\macplaceidx+1}
  + \maccovlen(\macplaceidx)$.\qed
\end{proof}
\end{full}

The following lemma gives an upper bound on $\maccovlen(\macplaceidx)$
using the recurrence relation obtained above.
\begin{lemma}
  \label{lem:covLenUpBound}
  $\maccovlen(\macplaceidx)\le
  (2\macnumplaces\macmaxarcw\macmaxcov\macmaxcov')^{\macnumplaces(\macplaceidx+1)!}$.
\end{lemma}
\begin{full}
\begin{proof}
  By induction on $\macplaceidx$. For $\macplaceidx=0$,
  $\maccovlen(0)\le \macnumplaces\macmaxcov\le
  (2\macnumplaces\macmaxarcw\macmaxcov\macmaxcov')^{\macnumplaces1!}$.

  $i=1$:
  \begin{align*}
    \maccovlen(1) &\le
    \macmaxcov'^{\macnumplaces}(\macmaxarcw\maccovlen(0)+\macmaxcov)+\maccovlen(0)\\
    & \le
    \macmaxcov'^{\macnumplaces}(\macmaxarcw\macnumplaces\macmaxcov+\macmaxcov)+
    \macnumplaces\macmaxcov\\
    & \le
    (\macmaxarcw\macmaxcov\macmaxcov')^{\macnumplaces}\macnumplaces\macmaxcov
    + \macnumplaces\macmaxcov \\
    & \le
    (\macnumplaces\macmaxarcw\macmaxcov\macmaxcov')^{2\macnumplaces} +
    \macnumplaces\macmaxcov \\
    &\le 
    2(\macnumplaces\macmaxarcw\macmaxcov\macmaxcov')^{2\macnumplaces}\\
    &\le
    (2\macnumplaces\macmaxarcw\macmaxcov\macmaxcov')^{\macnumplaces2!}
  \end{align*}
$i\ge 2$:
\begin{align*}
  \maccovlen(\macplaceidx+1) & \le
  \macmaxcov'^{\macnumplaces}(\macmaxarcw\maccovlen(\macplaceidx)+\macmaxcov)^{\macplaceidx+1}
  + \maccovlen(\macplaceidx)\\
  &\le
  \macmaxcov'^{\macnumplaces}(\macmaxarcw(2\macnumplaces\macmaxarcw\macmaxcov
  \macmaxcov')^{\macnumplaces(\macplaceidx+1)!}+\macmaxcov)^{\macplaceidx+1}
  + (2\macnumplaces\macmaxarcw\macmaxcov
  \macmaxcov')^{\macnumplaces(\macplaceidx+1)!}\\
  &\le
  (\macmaxarcw\macmaxcov\macmaxcov')^{\macnumplaces(\macplaceidx+1)}
  (2\macnumplaces\macmaxarcw\macmaxcov\macmaxcov')^{\macnumplaces(\macplaceidx+1)!
  (\macplaceidx+1)} + (2\macnumplaces \macmaxarcw \macmaxcov
  \macmaxcov')^{\macnumplaces (\macplaceidx+1)!}\\
  & \le (2\macnumplaces \macmaxarcw \macmaxcov
  \macmaxcov')^{\macnumplaces(\macplaceidx+1)}
  (2\macnumplaces\macmaxarcw\macmaxcov\macmaxcov')^{\macnumplaces(\macplaceidx+1)!
  (\macplaceidx+1)} + (2\macnumplaces \macmaxarcw \macmaxcov
  \macmaxcov')^{\macnumplaces (\macplaceidx+1)!}\\
  &\le (2\macnumplaces\macmaxarcw \macmaxcov\macmaxcov')^{
  \macnumplaces(\macplaceidx+1)( (\macplaceidx+1)!+1 )}+
  (2\macnumplaces \macmaxarcw \macmaxcov \macmaxcov')^{\macnumplaces
  (\macplaceidx+1)!}\\
  & \le 2(2\macnumplaces\macmaxarcw \macmaxcov\macmaxcov')^{
  \macnumplaces(\macplaceidx+1)( (\macplaceidx+1)!+1 )}\\
  &\le (2\macnumplaces\macmaxarcw \macmaxcov\macmaxcov')^{
  \macnumplaces(\macplaceidx+1)( (\macplaceidx+1)!+2 )}\\
  &\le (2\macnumplaces\macmaxarcw \macmaxcov\macmaxcov')^{
  \macnumplaces(\macplaceidx+2)!}
\end{align*}
The last step follows since
\begin{align*}
  \macplaceidx\ge 2 & \Rightarrow \macplaceidx!\ge 2\\
  & \Rightarrow (\macplaceidx+1)\macplaceidx!\ge 2(\macplaceidx+1)\\
  & \Rightarrow (\macplaceidx+1)! \ge 2(\macplaceidx+1)\\
  & \Rightarrow (\macplaceidx+1)(\macplaceidx+1)!+(\macplaceidx+1)!
  \ge (\macplaceidx+1)(\macplaceidx+1)!+2(\macplaceidx+1)\\
  &\Rightarrow (\macplaceidx+2)(\macplaceidx+1)! \ge
  (\macplaceidx+1)( (\macplaceidx+1)!+2 )\\
  &\Rightarrow (\macplaceidx+2)!\ge (\macplaceidx+1)(
  (\macplaceidx+1)!+2 )
\end{align*}\qed
\end{proof}
\end{full}

\begin{theorem}
  \label{thm:covInParaPspace}
  With the vertex cover number $\macvcnum$ and maximum arc weight
  $\macmaxarcw$ as parameters, the Petri net coverability problem can
  be solved in \macparapsp.
\end{theorem}
\begin{proof}
  From the \lemref{lem:covLenUpBound}, we get $\maccovlen(\macvcnum')\le
  (2\macnumplaces \macmaxarcw \macmaxcov \macmaxcov')^{ \macnumplaces
  (\macvcnum'+1)!}$. To guess and verify a covering sequence of length
  at most $\maccovlen(\macvcnum')$, a non-deterministic Turing machine
  needs to maintain a counter and intermediate markings, which can be
  done using memory size $\macOh(\macnumplaces (\macvcnum'+1)! (
  \macnumplaces \log |\macmark_{0}|+ \log\macnumplaces +
  \log\macmaxarcw + \log\macmaxcov + \log\macmaxcov'))$. An
  application of Savitch's theorem then gives us the \macparapsp{}
  algorithm. \qed
\end{proof}

\section{The boundedness problem}
\label{sec:boundedness}
In this section, we will show that with vertex cover number and
maximum arc weight as parameters, the Petri net boundedness problem
can be solved in \macparapsp{}. If there is a firing sequence
$\macfirseqo$ such that
$\macmark_{0}\macStep{\macfirseqo}\macmark_{1}$ and an intermediate
marking $\macmark$ such that $\macmark<\macmark_{1}$ (i.e.,
  $\macmark\le \macmark_{1}$ and
  $\macmark\ne \macmark_{1}$), then
$\macfirseqo$ is called a \macpopterm{self-covering sequence}. It is
well known that a Petri net is unbounded iff the initial marking
enables a self-covering sequence. Similar to the recurrence relation
for the length of covering sequences, Rackoff gave a recurrence relation
for the length of self-covering sequences also in \cite{RCK78}. We will
again use truncation lemma to prove that this recurrence relation
grows slowly for Petri nets with small vertex cover. The following
lemma formalizes the way truncation lemma is used in boundedness.
\begin{definition}
  \label{def:qEnabledSCS}
  Let $\macplacesu\subseteq \macplaces$ be a subset of places with
  $\macidplaces\subseteq \macplacesu$. Let $\macmark_{0}:\macplaces\to
  \macint$ be some function. A firing sequence
  $\macfirseqo=\mactranso_{1}\mactranso_{2}\cdots\mactranso_{\macfslidx}$
  is said to be a \macstandout{$\macplacesu$-enabled self-covering
  sequence} if there are intermediate functions
  $\macmark_{1},\macmark_{2},\dots,\macmark_{\macfslidx'},\dots,\macmark_{\macfslidx}$
  with $\macfslidx'<\macfslidx$ such that
  $\macmark_{0}\macstep{\mactranso_{1}}{\macplacesu} \macmark_{1}
  \macstep{\mactranso_{2}}{\macplacesu}\cdots
  \macstep{\mactranso_{\macfslidx'}}{\macplacesu}
  \macmark_{\macfslidx'} \macstep{}{} \cdots
  \macstep{\mactranso_{\macfslidx}}{\macplacesu}\macmark_{\macfslidx}$
  and $\macmark_{\macfslidx'}<\macmark_{\macfslidx}$. We call the
  subsequence between $\macmark_{\macfslidx'}$ and
  $\macmark_{\macfslidx}$ as the \macstandout{pumping portion} of the
  self-covering sequence.
\end{definition}
\begin{lemma}
  \label{lem:shortSCS}
  Suppose $\macplacesu\subseteq\macplaces$ is a subset of places with
  $\macidplaces\subseteq \macplacesu$. Let $\macmaxinit$ be the
  maximum of the range of the initial marking. If there is a
  $\macplacesu$-enabled self-covering sequence, then there is a
  $\macplacesu$-enabled self-covering sequence in which none of the
  places in $\macidplaces$ will have more than
  $\macmaxinit+\macmaxarcw+ \macmaxarcw^{2}+\macmaxarcw^{3}$ tokens in
  any intermediate marking.
\end{lemma}
\begin{full}
\begin{proof}
  Let
  $\macfirseqo=\mactranso_{1}\mactranso_{2}\cdots\mactranso_{\macfslidx}$
  be the $\macplacesu$-enabled self-covering sequence with
  $\macmark_{0}\macstep{\mactranso_{1}}{\macplacesu} \macmark_{1}
  \macstep{\mactranso_{2}}{\macplacesu}\cdots
  \macstep{\mactranso_{\macfslidx'}}{\macplacesu}
  \macmark_{\macfslidx'} \macstep{}{} \cdots
  \macstep{\mactranso_{\macfslidx}}{\macplacesu}\macmark_{\macfslidx}$
  and $\macmark_{\macfslidx'}<\macmark_{\macfslidx}$. First ensure
  that for every place $\macplaceo$ with
  $\macmark_{\macfslidx}(\macplaceo)>\macmark_{\macfslidx'}(\macplaceo)$,
  $\macmark_{\macfslidx}(\macplaceo)\ge
  \macmark_{\macfslidx'}(\macplaceo)+2\macmaxarcw$. If this is not the
  case, we can repeat the pumping portion of $\macfirseqo$
  $2\macmaxarcw$ times to ensure it. After this modification, let
  $\macfirseqo_{1}\macfirseqo_{2}$ be the $\macplacesu$-enabled
  self-covering sequence with $\macfirseqo_{2}$ being the pumping
  portion. Consider the $\macplacesu$-enabled self covering sequence
  $\macfirseqo_{1}\macfirseqo_{2}\macfirseqo_{2}$. For convenience, we
  will denote this sequence by $\macfirseqt_{1}\macfirseqt_{2}$, where
  $\macfirseqt_{1}=\macfirseqo_{1}\macfirseqo_{2}$ and
  $\macfirseqt_{2}=\macfirseqo_{2}$, with $\macfirseqt_{2}$ being the
  pumping portion.

  Consider a place $\macplaceo$ of some variety $\macplvar$ in
  $\macidplaces$. Let $\macmark$ be the last intermediate marking
  during the firing of $\macfirseqt_{1}$ from $\macmark_{0}$ such that
  $\macmark(\macplaceo)$ is the minimum number of tokens in
  $\macplaceo$ among all intermediate markings.

  \emph{Case 1:} $\macmark(\macplaceo) \ge \macmark_{0}(\macplaceo)$. In
  this case, the number of tokens in $\macplaceo$ does not come below
  $\macmark_{0}(\macplaceo)$ at all. Let $\macfirseqt'$ be the
  sub-word of $\macfirseqt_{1}\macfirseqt_{2}$ consisting of all
  transitions having an arc to/from $\macplaceo$. Transfer $\macfirseqt'$
  to $\macplaceo_{\macplvar}$. If the number of tokens in $\macplaceo$
  was being increased by $\macfirseqt_{2}$ before the transfer, the
  transfer will result in the number of tokens in $\macplaceo$
  remaining unchanged during the pumping portion. To remedy this,
  identify the last transition that adds tokens to
  $\macplaceo_{\macplvar}$ and transfer it back to $\macplaceo$. Since
  $\macfirseqt_{2}$ was adding at least $2\macmaxarcw$ tokens to
  $\macplaceo_{\macplvar}$ (which we ensured in the beginning of this
  proof), the above mentioned transfer of one transition back to
  $\macplaceo$ will not affect firability of any transition and will
  also ensure that the number of tokens in both $\macplaceo$ and
  $\macplaceo_{\macplvar}$ increase during pumping portion
  $\macfirseqt_{2}$.

  \emph{Case 2:} $\macmark(\macplaceo) < \macmark_{0}(\macplaceo)$.
  Invoking truncation lemma with $\macnumtok=\macmark_{0}(\macplaceo)+
  \macmaxarcw$, we identify sub-words between $\macmark_{0}$ and
  $\macmark$ and transfer them to $\macplaceo_{\macplvar}$ so that in
  any intermediate marking, $\macplaceo$ has at most
  $\macmaxinit+\macmaxarcw+\macmaxarcw^{2}+\macmaxarcw^{3}$ tokens.
  Let $\macfirseqt'$ be the sub-word of
  $\macfirseqt_{1}\macfirseqt_{2}$ consisting all transitions
  having an arc to/from $\macplaceo$, occurring between $\macmark$ and the
  final marking reached. This sub-word $\macfirseqt'$ is safe for
  transfer from $\macplaceo$ to $\macplaceo_{\macplvar}$ (since
  $\macmark(\macplaceo)$ is the minimum number of tokens in
  $\macplaceo$ reached during the firing of $\macfirseqt_{1}$ and
  $\macfirseqt_{2}$ will not decrease the number of tokens in
  $\macplaceo$ below $\macmark(\macplaceo)$ in any intermediate
  marking after $\macmark$) and we transfer it to
  $\macplaceo_{\macplvar}$. Again, if $\macfirseqt_{2}$ was increasing
  the number of tokens in $\macplaceo$ before the above transfer,
  identify the last transition adding tokens to
  $\macplaceo_{\macplvar}$ and transfer it back to $\macplaceo$. As in
  the first case, this will ensure that the number of tokens in both
  $\macplaceo$ and $\macplaceo_{\macplvar}$ increase during pumping
  portion $\macfirseqt_{2}$.

  For every independent place $\macplaceo\in \macidplaces$, we
  identify and transfer sub-words to $\macplaceo_{\macplvar}$ based
  on one of the above two cases. Finally, we end up with a
  $\macplacesu$-enabled self-covering sequence in which none of the
  independent places will have more than $\macmaxinit + \macmaxarcw +
  \macmaxarcw^{2} + \macmaxarcw^{3}$ tokens in any intermediate
  marking.\qed
\end{proof}
\end{full}

Before we can use \lemref{lem:shortSCS}, we need the following
technical lemmas. The first one is an adaptation of Lemma 4.5 in
Rackoff's paper \cite{RCK78} to our setting.
\begin{lemma}
  \label{lem:smallSolnShortSCS}
  Let $\macplacesu\subseteq \macplaces$ with $\macidplaces\subseteq
  \macplacesu$ and $\macmaxinit' \in \macnat$ be such that there is a
  $\macplacesu$-enabled self-covering sequence from some
  $\macmark_{0}$ in which all intermediate markings have at most
  $\macmaxinit'$ tokens in any independent place. Also suppose that
  all intermediate markings have at most $\macnumtok$ tokens in any
  place in $\macplacesu \setminus \macidplaces$. Then, there is a
  $\macplacesu$-enabled self-covering sequence of length at most
  $8\macvcnum'(2\macnumtok)^{\macicon'\macvcnum'^{3}}(\macmaxinit'
  \macmaxarcw)^{\macicon'\macnumplaces^{4}}$ for some constant
  $\macicon'$.
\end{lemma}
\begin{full}
\begin{proof}
  Suppose the given self-covering sequence is of the form
  $\macmark_{0}\macstep{\macfirseqo_{1}}{\macplacesu}\macmark_{1}
  \macstep{\macfirseqo_{2}}{\macplacesu} \macmark_{2}$ with
  $\macfirseqo_{2}$ being the pumping portion. The length of
  $\macfirseqo_{1}$ is at most
  $\macmaxinit'^{\macnumplaces}\macnumtok^{\macvcnum'}$. For reducing
  the length of $\macfirseqo_{2}$, we will closely follow the proof of
  Lemma 4.5 in Rackoff's paper \cite{RCK78}. Let a $\macplacesu$-loop
  be any sequence of transitions whose total effect is $0$ on any
  place in $\macplacesu$.

  As in Rackoff's proof of Lemma 4.5 in \cite{RCK78}, remove
  $\macplacesu$-loops from $\macfirseqo_{2}$ carefully until what
  remains behind is a sequence $\macfirseqo_{2}'$ of length at most
  $(\macmaxinit'^{\macnumplaces}\macnumtok^{\macvcnum'}+1)^{2}$. Let
  $\macpupcv\in \macnat^{\macvcnum'}$ be a vector containing a
  $1$ in each coordinate corresponding to a special place in
  $\macspplaces$ whose number of tokens is increased by
  $\macfirseqo_{2}$ and $0$ in all other coordinates. If $\macfirseqt$
  is a $\macplacesu$-loop, its \macstandout{loop value} is the vector
  in $\macint^{\macvcnum'}$, which contains in each coordinate the
  total effect of $\macfirseqt$ on the corresponding special place in
  $\macspplaces$. Let $\maclvs\subseteq \macint^{\macvcnum'}$ be the
  set of loop values that were removed from $\macfirseqo_{2}$. Let
  $\maclvm$ be the matrix with $\macvcnum'$ rows, whose columns are
  the members of $\maclvs$. For any sequence $\macfirseqt$, let
  $\macef(\macfirseqt)$ be the vector in $\macint^{\macvcnum'}$, which
  contains in each coordinate the total effect of $\macfirseqt$ on the
  corresponding special place in $\macspplaces$. Since
  $\macfirseqo_{2}$ is a pumping portion,
  $\macef(\macfirseqo_{2})\ge \macpupcv$. Now, the effect of
  $\macfirseqo_{2}$ can be split into the effect of
  $\macfirseqo_{2}'$ and the effect of $\macplacesu$-loops that were
  removed from $\macfirseqo_{2}$. If $\macsolv(\maclvidx)$ is the
  number of $\macplacesu$-loops removed from $\macfirseqo_{2}$ whose
  loop value is equal to the $\maclvidx$\textsuperscript{th} column of
  $\maclvm$, then we have $\maclvm\macsolv\ge
  \macpupcv-\macef(\macfirseqo_{2}')$.

  A loop value is just the effect of at most
  $\macnumtok^{\macvcnum'}\macmaxinit'^{\macnumplaces}$ transitions,
  and hence each entry of $\maclvm$ is of absolute value at most
  $\macnumtok^{\macvcnum'}\macmaxinit'^{\macnumplaces}\macmaxarcw$.
  The matrix $\maclvm$ has therefore at most
  $(2\macnumtok^{\macvcnum'}\macmaxinit'^{\macnumplaces}\macmaxarcw+1)^{\macvcnum'}$
  columns. Each entry of $\macpupcv-\macef(\macfirseqo_{2}')$ is of
  absolute value at most
  $\macmaxarcw(\macnumtok^{\macvcnum'}\macmaxinit'^{\macnumplaces}+1)^{2}+1$.
  Letting $d_{1}=\macvcnum'$ and $d=\max\{
  (2\macnumtok^{\macvcnum'}\macmaxinit'^{\macnumplaces}\macmaxarcw+1)^{\macvcnum'},
  \macnumtok^{\macvcnum'}\macmaxinit'^{\macnumplaces}\macmaxarcw,
  \macmaxarcw(\macnumtok^{\macvcnum'}\macmaxinit'^{\macnumplaces}+1)^{2}+1\}\le
  (2\macnumtok)^{3\macvcnum'}(\macmaxinit'\macmaxarcw)^{3\macnumplaces^{2}}$,
  we can apply Lemma 4.4 of \cite{RCK78}. The result is that there is
  a vector $\macssolv\in \macnat^{|\maclvs|}$ such that the sum of
  entries of $\macssolv$ is equal to $l_{1}\le d(
  (2\macnumtok)^{3\macvcnum'}
  (\macmaxinit'\macmaxarcw)^{3\macnumplaces^{2}}
  )^{\macicon\macvcnum'}$ for some constant $\macicon$. Let
  $\macicon'$ be a constant such that $l_{1}\le \macvcnum'
  (2\macnumtok)^{\macicon'\macvcnum'^{2}}
  (\macmaxinit'\macmaxarcw)^{\macicon'\macnumplaces^{3}}$.

  Now, we will put $l_{1}$ $\macplacesu$-loops back to
  $\macfirseqo_{2}'$, which was of length at most
  $(\macnumtok^{\macvcnum'}\macmaxinit'^{\macnumplaces}+1)^{2}$. Since
  the length of each $\macplacesu$-loop is at most
  $\macnumtok^{\macvcnum'}\macmaxinit'^{\macnumplaces}$, the total
  length of the newly constructed pumping portion is at most
  $(\macnumtok^{\macvcnum'}\macmaxinit'^{\macnumplaces}+1)^{2}+\macvcnum'
  (2\macnumtok)^{\macicon'\macvcnum'^{3}}(\macmaxinit'\macmaxarcw)^{\macicon'\macnumplaces^{4}}$.
  Together with $\macfirseqo_{1}$, whose length is at most
  $\macnumtok^{\macvcnum'}\macmaxinit'^{\macnumplaces}$, we get a
  $\macplacesu$-enabled self-covering sequence of length at most
  $2(\macnumtok^{\macvcnum'}\macmaxinit'^{\macnumplaces}+1)^{2}+\macvcnum'
  (2\macnumtok)^{\macicon'\macvcnum'^{3}}(\macmaxinit'\macmaxarcw)^{\macicon'\macnumplaces^{4}}
  \le 8\macvcnum'
  (2\macnumtok)^{\macicon'\macvcnum'^{3}}(\macmaxinit'\macmaxarcw)^{\macicon'\macnumplaces^{4}}$.\qed
\end{proof}
\end{full}

\begin{definition}
  \label{def:shortestScs}
   Let $\macmaxinit'\in \macnat$ be some fixed number (we will later use
  it to denote $\macmaxinit+\macmaxarcw+\macmaxarcw^{2} +
  \macmaxarcw^{3}$, as in \lemref{lem:shortSCS}). For
  $\macidnidx\in \macnat$, $\macplacesu\subseteq \macplaces$ with $\macidplaces\subseteq
  \macplacesu$ and a function $\macmark:\macplaces\to \macint$, let
  $\macslencov(\macplacesu,\macidnidx,\macmark)$ be the length of the
  shortest $\macplacesu$-enabled self-covering sequence from $\macmark$
  if there is a $\macplacesu$-enabled self-covering sequence from
  $\macmark$ in which all intermediate markings have at most
  $\macmaxinit'+\macidnidx\macmaxarcw$ tokens in any independent place.
  Let $\macslencov(\macplacesu,\macidnidx,\macmark)$ be $0$ if there is
  no such sequence. Define $\macscovlen(\macplaceidx,\macidnidx)=
  \max\left\{ \macslencov(\macplacesu,\macidnidx,\macmark)\mid
  \macidplaces\subseteq \macplacesu \subseteq \macplaces,
  |\macplacesu\setminus\macidplaces|=\macplaceidx,\macmark:\macplaces\to
  \macint\right\}$. 
\end{definition}
The following lemma is an immediate consequence of Lemma 4.5 in
\cite{RCK78}.
\begin{lemma}
  \label{lem:scsBaseCase}
  There is a constant $\macicont$ such that
  $\macscovlen(0,\macidnidx)\le
  (\macmaxinit'+\macidnidx\macmaxarcw)^{\macnumplaces^{\macicont}}$.
\end{lemma}

\begin{lemma}
  \label{lem:scsIndStep}
  $\macscovlen(\macplaceidx+1,\macidnidx)\le 8\macvcnum'(2\macmaxarcw
  \macscovlen(\macplaceidx,\macidnidx+1))^{\macicon\macvcnum'^{3}}
  ( (\macmaxinit'+\macidnidx\macmaxarcw)\macmaxarcw
  )^{\macicon'\macnumplaces^{4}}$ for some appropriately chosen
  constants $\macicon$ and $\macicon'$.
\end{lemma}
\begin{full}
\begin{proof}
  Suppose $\macplacesu\subseteq \macplaces$ such that
  $\macidplaces\subseteq \macplacesu$ and $|\macplacesu\setminus
  \macidplaces|=\macplaceidx+1$. Also suppose that there is a
  $\macplacesu$-enabled self-covering sequence from some marking
  $\macmark$ such that all intermediate markings have at most
  $\macmaxinit'+\macidnidx\macmaxarcw$ tokens in any independent
  place. If all intermediate markings have at most $\macmaxarcw
  \macscovlen(\macplaceidx,\macidnidx+1)$ tokens in any place in
  $\macplacesu\setminus \macidplaces$, the required result is a
  consequence of \lemref{lem:smallSolnShortSCS}, substituting $\macmaxarcw
  \macscovlen(\macplaceidx,\macidnidx+1)$ for $\macnumtok$ and
  $\macmaxinit'+\macidnidx\macmaxarcw$ for $\macmaxinit'$.

  Otherwise, let $\macfirseqo=\macfirseqo_{1}\macfirseqo_{2}$ be the
  self-covering sequence, with $\macfirseqo_{2}$ being the pumping
  portion. Ensure that for any independent place $\macplaceo$,
  $\macfirseqo_{2}$ adds at most $\macmaxarcw$ tokens (otherwise, we
  can transfer from $\macplaceo$ to $\macplaceo_{\macplvar}$ the last
  transition that adds tokens to $\macplaceo$, where $\macplvar$ is
  the variety of $\macplaceo$). Let $\macmark_{1}$ be the first
  intermediate marking with more than
  $\macmaxarcw\macscovlen(\macplaceidx,\macidnidx+1)$ tokens in some
  special place $\macplacet\in \macplacesu\setminus\macidplaces$. Let
  the subsequence up to $\macmark_{1}$ be called $\macfirseqt_{1}$ and
  rest of the sequence be called $\macfirseqt_{2}$ (the pumping
  portion $\macfirseqo_{2}$ is a suffix of
  $\macfirseqo=\macfirseqt_{1}\macfirseqt_{2}$). The length of
  $\macfirseqt_{1}$ is at most $(\macmaxarcw\macscovlen(\macplaceidx,
  \macidnidx+1))^{\macvcnum'}(\macmaxinit'+\macidnidx\macmaxarcw)^{\macnumplaces}$.
  Starting from $\macmark_{1}$, $\macfirseqt_{2}\macfirseqo_{2}$ is a
  $\macplacesu$-enabled self-covering sequence. At the end of
  $\macfirseqt_{2}$, every independent place has at most
  $\macmaxinit'+\macidnidx\macmaxarcw$ tokens. During the firing of
  $\macfirseqo_{2}$ after $\macfirseqt_{2}$, every independent place
  has at most $\macmaxinit'+ (\macidnidx+1)\macmaxarcw$ tokens in any
  intermediate marking (since $\macfirseqo_{2}$ adds at most
  $\macmaxarcw$ tokens to every independent place; see
  \figref{fig:ScsIndStepProof}).
  \begin{figure}[!htp]
    \begin{center}
      \includegraphics{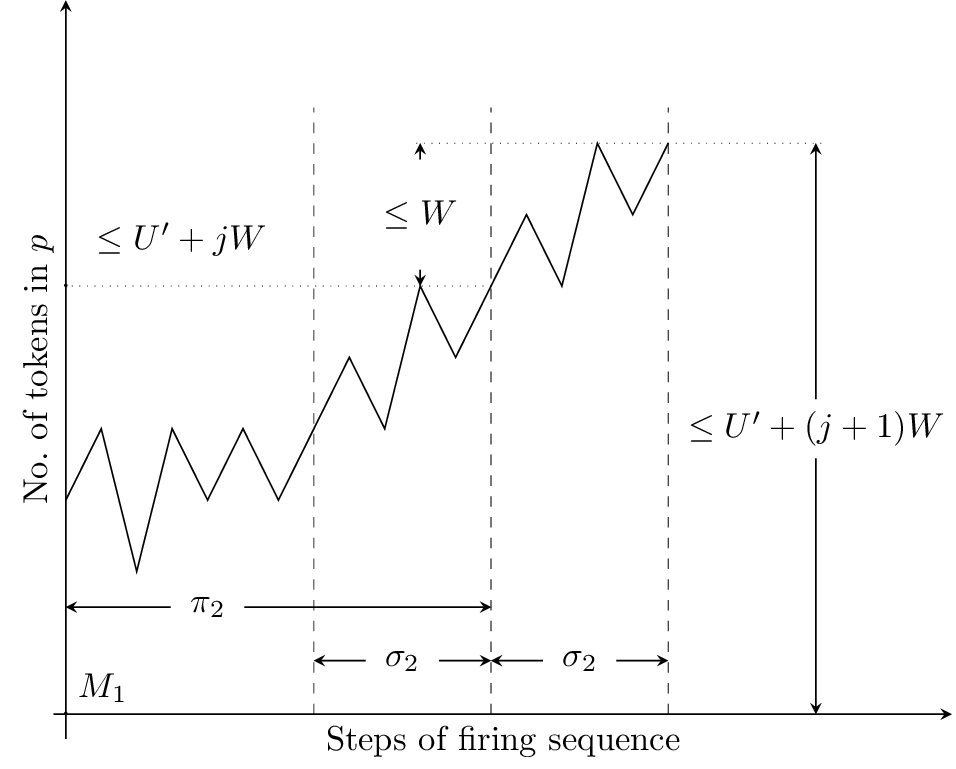}
    \end{center}
    \caption{Illustration for proof of \lemref{lem:scsIndStep}}
    \label{fig:ScsIndStepProof}
  \end{figure}

  Hence, $\macfirseqt_{2}\macfirseqo_{2}$ is a $\macplacesu\setminus
  \left\{ \macplacet \right\}$-enabled self-covering sequence from
  $\macmark_{1}$ such that in all intermediate markings, every
  independent place has at most
  $\macmaxinit'+(\macidnidx+1)\macmaxarcw$ tokens. By definition,
  there is a $\macplacesu\setminus \{\macplacet\}$-enabled
  self-covering sequence $\macfirseqt_{2}'$ from $\macmark_{1}$ of
  length at most $\macscovlen(\macplaceidx,\macidnidx+1)$. Since
  $\macmark_{1}(\macplacet)\ge
  \macmaxarcw\macscovlen(\macplaceidx,\macidnidx+1)$ and
  $\macmark\macstep{\macfirseqt_{1}}{\macplacesu}\macmark_{1}$,
  $\macfirseqt_{1}\macfirseqt_{2}'$ is a $\macplacesu$-enabled
  self-covering sequence from $\macmark$ of length at most
  $(\macmaxarcw\macscovlen(\macplaceidx,
  \macidnidx+1))^{\macvcnum'}(\macmaxinit'+\macidnidx\macmaxarcw)^{\macnumplaces}
  + \macscovlen(\macplaceidx,\macidnidx+1)$ .\qed
\end{proof}
\end{full}

Now using \lemref{lem:shortSCS}, we can conclude that
if there is a self-covering sequence, there is one of length at most
$\macscovlen(\macvcnum',1)$, setting $\macmaxinit'=\macmaxinit+
\macmaxarcw^{2}+ \macmaxarcw^{3}$ in the definition of $\macscovlen$.
The following lemma gives an upper bound on this quantity. We use
$\maciconh$ to denote $\macicon'\macvcnum'^{3}$.
\begin{lemma}
  \label{lem:scovlenupbound}
  $\macscovlen(\macplaceidx,\macidnidx)\le
  (8\macvcnum')^{(1+\maciconh)^{\macplaceidx}}
  (2\macmaxarcw)^{\macpoly_{1}(\maciconh^{\macplaceidx})}
  (\macmaxinit'+(\macidnidx+\macplaceidx)\macmaxarcw)^{\macpoly_{2}(\maciconh^{\macplaceidx})}$
  where $\macpoly_{1}(\maciconh^{\macplaceidx})$ and
  $\macpoly_{2}(\maciconh^{\macplaceidx})$ are polynomials in
  $\maciconh^{\macplaceidx},\macicon',\macvcnum'$ and $\macnumplaces$.
\end{lemma}
\begin{full}
\begin{proof}
  By induction on $\macplaceidx$. $\macscovlen(0,\macidnidx)\le
  (\macmaxinit'+\macidnidx\macmaxarcw)^{\macnumplaces^{\macicont}} \le
  8\macvcnum'(\macmaxinit'+\macidnidx\macmaxarcw)^{\macnumplaces^{\macicont}}$.
  \begin{align*}
    \macscovlen(\macplaceidx+1,\macidnidx) &\le 8\macvcnum'
    (2\macmaxarcw\macscovlen(\macplaceidx,\macidnidx+1))^{\maciconh}
    ( (\macmaxinit'+\macidnidx\macmaxarcw)\macmaxarcw
    )^{\macicon'\macnumplaces^{4}}\\
    &\le 8\macvcnum'\left[2\macmaxarcw
    (8\macvcnum')^{(1+\maciconh)^{\macplaceidx}}
    (2\macmaxarcw)^{\macpoly_{1}(\maciconh^{\macplaceidx})}
    (\macmaxinit'+(\macidnidx+1+\macplaceidx)\macmaxarcw)^{\macpoly_{2}(\maciconh^{\macplaceidx})}\right]^{\maciconh}\\
    &\quad( (\macmaxinit'+\macidnidx\macmaxarcw)\macmaxarcw
    )^{\macicon'\macnumplaces^{4}}\\
    &\le (8\macvcnum')^{1+\maciconh(1+\maciconh)^{\macplaceidx}}
    (2\macmaxarcw)^{(1+\macpoly_{1}(\maciconh^{\macplaceidx}))\maciconh+\macicon'\macnumplaces^{4}}
    (\macmaxinit'+(\macidnidx+\macplaceidx+1)\macmaxarcw)^{\macpoly_{2}(\maciconh^{\macplaceidx})\maciconh
    +\macicon'\macnumplaces^{4}}
  \end{align*}
  It is now enough to choose $\macpoly_{1}$ and $\macpoly_{2}$ such
  that $\macpoly_{1}(\maciconh^{\macplaceidx+1})\ge
  (1+\macpoly_{1}(\maciconh^{\macplaceidx}))\maciconh+\macicon'\macnumplaces^{4}$,
  $\macpoly_{2} (\maciconh^{0})\ge \macnumplaces^{\macicont}$ and
  $\macpoly_{2}(\maciconh^{\macplaceidx+1})\ge
  \macpoly_{2}(\maciconh^{\macplaceidx})\maciconh +
  \macicon'\macnumplaces^{4}$. These conditions are met by
  $\macpoly_{1}(\maciconh^{\macplaceidx})=(\maciconh+\macicon'\macnumplaces^{4})(\maciconh^{\macplaceidx}-1)$
  and
  $\macpoly_{2}(\maciconh^{\macplaceidx})=\maciconh^{\macplaceidx}
  \macnumplaces^{\macicont} + \macicon'\macnumplaces^{4}(
  \maciconh^{\macplaceidx}-1)$, assuming $\maciconh\ge 2$.\qed
\end{proof}
\end{full}
\begin{theorem}
  \label{thm:boundInParaPsp}
  With the vertex cover number $\macvcnum$ and maximum arc weight
  $\macmaxarcw$ as parameters, the Petri net boundedness problem can
  be solved in \macparapsp.
\end{theorem}
\begin{proof}
  A non-deterministic Turing machine can test for unboundedness by
  guessing and verifying the presence of a self-covering sequence of
  length at most $\macscovlen(\macvcnum',1)$. By
  \lemref{lem:scovlenupbound}, the memory needed by such a Turing
  machine is bounded by $\macOh(\macnumplaces \log|\macmark_{0}|+ \macnumplaces + \log \macmaxarcw +
  (1+\macicon'\macvcnum'^{3})^{\macvcnum'}\log \macvcnum' +
  \macpoly_{1}(\macicon'^{\macvcnum'}\macvcnum'^{3\macvcnum'})\log
  \macmaxarcw +
  \macpoly_{2}(\macicon'^{\macvcnum'}\macvcnum'^{3\macvcnum'})
  \log(\macmaxinit'\macvcnum'\macmaxarcw))$, or
  $\macOh(\macnumplaces \log|\macmark_{0}| + \macnumplaces +
  \macpoly(\macicon'^{3\macvcnum'} \macvcnum'^{3\macvcnum'})\log(\macmaxinit'\macvcnum'\macmaxarcw))$
  for some polynomial $\macpoly$. An application of Savitch's theorem
  now gives us the \macparapsp{} algorithm for boundedness. \qed
\end{proof}

\section{A logic based on Coverability and Boundedness}
\label{sec:logic}
Following is a logic (borrowed from \cite{PL09}) of
properties such that its model checking can be reduced to 
coverability ($\kappa$) and boundedness ($\beta$) problems, but is designed
to avoid expressing reachability. This is a fragment of Computational
Tree Logic (CTL).
\begin{align*}
\tau &::= p,~p\in P~|~\tau_1 + \tau_2~|~c\tau,~c \in \macnat\\
\kappa &::= \tau\ge c,~c\in \macnat~|~
	\kappa_1\land\kappa_2~|~\kappa_1\lor\kappa_2~|~\EF\kappa\\
\beta &::= \{\tau_{1},\dots,\tau_{r}\}\bdd~|~\lnot\beta~|~\beta_1\lor\beta_2\\
\phi &::= \beta~|~\kappa~|~\phi_1 \land \phi_2~|~\phi_1 \lor \phi_2
\end{align*}

\begin{conf}
  The semantics of the above logic is explained in the full version
  with examples.
\end{conf}
\begin{full}
The satisfaction of a formula $\phi$ by a Petri net $\macnet$ with
initial marking $M_{0}$ 
(denoted as $\macnet,M_{0}\models \phi$) is defined below. 
The boolean operators work as usual.
Note that every term (of type $\tau$)
gives a function $\linf_{\tau}:P\to \macnat$ such that $\tau$ is
syntactically equivalent to $\sum_{p\in P}\linf_{\tau}(p)p$.
\begin{itemize}
  \item $\macnet,M_{0}\models \tau\ge c$ if $\sum_{p\in
    P}\linf_{\tau}(p)M_{0}(p)\ge c$.
  \item $\macnet,M_{0}\models\EF\kappa$ if there is a marking
    $\macmark$ reachable from $\macmark_{0}$ such that $\macnet,M\models\kappa$.
  \item $\macnet,M_{0}\models\{\tau_{1},\dots,\tau_{r}\}\bdd$ if
    $\exists \macboundOnP\in \macnat$ such that for all markings
    $\macmark$ reachable from $\macmark_{0}$,  there is a $j\in
    \{1,\dots,r\}$ such that $\sum_{p\in P}L_{\tau_{j}}(p)M(p)\le
    \macboundOnP$.
\end{itemize}

In the Petri net of \figref{fig:pnExample}, if we set
$\macmark_{cov}$ as
$\macmark_{cov}(\macplaceo_{1})=\macmark_{cov}(\macplaceo_{2})=1$ and
$\macmark_{cov}(\macplaceo_{3})=0$, the coverability of
$\macmark_{cov}$ can be expressed as $\EF(\macplaceo_{1}\ge 1 \land
\macplaceo_{2}\ge 1)$. Boundedness of the Petri net in
\figref{fig:pnExample} can be expressed as
$\{\macplaceo_{1}+\macplaceo_{2} + \macplaceo_{3}\}\bdd$. If the
$\kappa$ formulas of the above logic had allowed formulas of type
$\tau\le c$, then we could have expressed reachability of
$\macmark_{cov}$ as $\EF(\macplaceo_{1}\ge 1 \land
\macplaceo_{1}\le 1 \land \macplaceo_{2}\ge 1 \land
\macplaceo_{2}\le 1 \land \macplaceo_{3} \le 0)$. Since much less is
known about the complexity of reachability, the above logic is
designed to avoid expressing reachability.
\end{full}

\begin{theorem}
  \label{thm:FormulaSatParaPspace}
  Given a Petri net with an initial marking and a formula $\phi$, if
  the vertex cover number $\macvcnum$ and the maximum arc weight
  $\macmaxarcw$ of the net are treated as parameters and the nesting
  depth $\nestDepth$ of $\EF$ modality in the formula is treated as a
  constant, then there is a \macparapsp{} algorithm that checks if the
  net satisfies the given formula.
\end{theorem}

\begin{full}
   The details of model checking $\kappa$ formulas is given in
   \subsecref{sec:kappaFormulas}.
\end{full}
\begin{conf}
  The details of model checking $\kappa$ formulas is given in
   the full version of this paper.
\end{conf}
While reading \cite{SD2010}, we realized that there is a mistake in
the reduction from model checking $\beta$ formulas to checking
the presence of self-covering sequences that we gave in \cite{PL09}.
However, it can be corrected using the notion of
\macpopterm{disjointness sequences} introduced by Demri in
\cite{SD2010}.
\begin{full}
  \Subsecref{sec:betaFormulas}
  gives the details of a \macparapsp{} algorithm for model checking
  $\beta$ formulas using ideas borrowed from \cite{SD2010}.
\end{full}
\begin{conf}
  The full version of this paper contains the details of a
  \macparapsp{} algorithm for model checking $\beta$ formulas using
  ideas borrowed from \cite{SD2010}.
\end{conf}

\begin{full}
\subsection{Model checking $\kappa$ formulas}
\label{sec:kappaFormulas}
We now consider verifying the formulas $\kappa$. We first reduce the
formulas to the form of $\gamma \land \EF( \kappa_{1}) \land \cdots
\land \EF( \kappa_{r})$, with $\gamma$ having only conjunctions of
$\tau \ge \macicon$ formulas by nondeterministically choosing
disjuncts from subformulas of $\kappa$. We call $\gamma$ the
\macstandout{content} of $\kappa$ and $\kappa_{1},\dots,\kappa_{r}$
the \macstandout{children} of $\kappa$. Each of the children may have
their own content and children, thus generating a tree with nodes
$\treenodes$, with $\kappa$ at the root of this tree.  We will
represent the nodes of this tree by sequences of natural numbers, $0$
being the root.

The maximum length of sequences in $\treenodes$ is one more than the
nesting depth of the $\EF$ modality in $\kappa$ and we denote it by
$\nestDepth$. Let $[\nestDepth] = \{0,1,\dots,\nestDepth-1\}$. 
If $\treenode\in \treenodes$ is a tree node that
represents the formula $\kappa(\treenode)=\gamma\land
\EF(\kappa_{1})\land\cdots\land\EF(\kappa_{r})$,
$\nodecontent(\treenode)=\gamma$ denotes the content of the node
$\treenode$. 
Let $ratio(\tau \ge c) = max\{\lceil c/L_{\tau}(p) \rceil \mid 
	\linf_{\tau}(p) \neq 0, p \in P\}$.
Defining $\max(\emptyset)=0$, we define
the maximum ratio at height $i$ in the tree by
$ratio(i) = \max\{ratio(\tau \ge c) \mid \tau\ge c
\text{ appears as a conjunct in }\nodecontent(\treenode)\text{ for some
}\treenode\in\treenodes,|\treenode|=i+1\}$.
\begin{definition}
  \label{def:BoundFunction}
  Recalling \defref{def:QCovSeq}, let $\maccovlen'(
  \macmark_{cov}) = \max\{ \maclencov(\macplaces, \macmark,
  \macmark_{cov}) \mid \macmark: \macplaces \to \macint\}$. Given a
  formula $\kappa$ and a Petri net $\macnet$ with initial marking
  $M_{0}$, the bound function $\boundfn:[\nestDepth]\times P\to
  \macnat$ is defined as follows.  We use $\boundfn(j)$ for the
  marking defined by
  $\boundfn(j)(p)=\boundfn(j,p)$.  \begin{itemize} \item[$\bullet$]
      $\boundfn(\nestDepth-1,p)=ratio(\nestDepth-1)$, \item[$\bullet$]
      $\boundfn(\nestDepth-i,p)=\max\{ratio(\nestDepth-i),
      \wt\maccovlen'(\boundfn(\nestDepth-i+1))+\boundfn(\nestDepth-i+1,p)\}$,
      $1<i<\nestDepth$, \item[$\bullet$] $\boundfn(0,p)=M_{0}(p)$.
  \end{itemize} A guess function $\guessfn:\treenodes\times P\to
  \macnat$ is any function that satisfies $\guessfn(\treenode,p)\le
  \boundfn(|\treenode|-1,p)$ for all $\treenode\in \treenodes$ and
  $p\in P$. If $\guessfn$ is a guess function, $\guessfn(\treenode)$
  is the marking defined by
  $\guessfn(\treenode)(p)=\guessfn(\treenode,p)$.
\end{definition}
If a given Petri net satisfies the formula $\kappa=\gamma\land
\EF(\kappa_{1})\land\cdots\land\EF(\kappa_{r})$, then there exist
firing sequences $\sigma_{01},\dots,\sigma_{0r}$ that are all enabled
at the initial marking $M_{0}$ such that
$M_{0}\macStep{\sigma_{0i}}M_{0i}$ and $M_{0i}$ satisfies $\kappa_{i}$. In
general, if $\kappa$ generates a tree with set of nodes $\treenodes$,
then there is a set of sequences $\{\sigma_{\treenode}\mid
\treenode\in \treenodes\setminus\{0\}\}$ and set of markings
$\{M_{\treenode}\mid \treenode\in \treenodes\}$ such that
$M_{\treenode}\macStep{\sigma_{\treenode j}}M_{\treenode j}$ for all
$\treenode,\treenode j\in \treenodes$ and $M_{\treenode}$ satisfies
$\nodecontent(\treenode)$ for all $\treenode \in \treenodes$.
\begin{lemma}
  \label{lem:GuessFunctionPNPath}
  There exist sequences $\{\mu_{\treenode}\mid \treenode\in
  \treenodes\setminus \{0\}\}$ and markings $\{M_{\treenode}\mid
  \treenode\in \treenodes\}$ such that
  $M_{\treenode}\macStep{\mu_{\treenode j}}M_{\treenode j}$ for all
  $\treenode,\treenode j\in \treenodes$ with $M_{\treenode}$
  satisfying $\nodecontent(\treenode)$ and $|\mu_{\treenode}|\le
  \maccovlen'(\boundfn(|\treenode|-1))$ iff there exist sequences
  $\{\sigma_{\treenode}\mid \treenode\in \treenodes\setminus \{0\}\}$
  and markings $\{M'_{\treenode}\mid \treenode\in \treenodes\}$
  ($M_{0}'$ should be equal to $M_{0}$) such
  that $M'_{\treenode}\macStep{\sigma_{\treenode j}}M'_{\treenode j}$ for all
  $\treenode,\treenode j\in \treenodes$ with $M'_{\treenode}$
  satisfying $\nodecontent(\treenode)$.
\end{lemma}
\begin{proof}
  ($\Rightarrow$) Since $M_{\treenode}$ satisfies
  $\nodecontent(\treenode)$, we can take
  $M_{\treenode}'=M_{\treenode}$ and $\sigma_{\treenode}=\mu_{\treenode}$.

  ($\Leftarrow$) Consider the following guess function:
  \begin{displaymath}
    \guessfn(\treenode,p)=\left\{ \begin{array}{ll}
      M_{0}(p) & \text{if } \treenode=0\\
      M_{\treenode}'(p) & \text{if }\treenode\ne 0\text{ and }M_{\treenode}'(p)\le
      \boundfn(|\treenode|-1,p)\\
      \boundfn(|\treenode|-1,p) & \text{otherwise}\end{array} \right.
  \end{displaymath}
  By definition, $\guessfn(\treenode)\le M_{\treenode}'$ and
  $\guessfn(\treenode)\le \boundfn(|\treenode|-1)$. Since
  $\sigma_{\treenode j}$ is a firing sequence that covers
  $M_{\treenode j}'$ from
  $M_{\treenode}'$, there exist sequences
  $\mu_{\treenode j}$ that cover $\guessfn(\treenode j)$ starting
  from $M_{\treenode}'$ whose length
  is at most $\maccovlen'(\guessfn(\treenode j))$ (and hence at most
  $\maccovlen'(\boundfn(|\treenode j|-1))$). We claim that
  there exist markings $\{M_{\treenode}\mid \treenode\in \treenodes\}$
  such that $M_{\treenode}\macStep{\mu_{\treenode j}}M_{\treenode j}$
  for all $\treenode,\treenode j\in \treenodes$ and that
  $M_{\treenode}$ satisfies $\nodecontent(\treenode)$ for all
  $\treenode \in \treenodes$.
  
  First, we claim that every $\mu_{\treenode j}$ can be fired from
  $M_{\treenode}$ and that every place $p$ will satisfy at least one of the
  following two conditions:
  \begin{enumerate}
    \item $M_{\treenode j}(p)\ge M_{\treenode j}'(p)$
    \item $M_{\treenode j}(p)\ge \boundfn(|\treenode j|-1,p)$
  \end{enumerate}
  We will prove this claim by induction on $|\treenode|$.

  \emph{Base case}: $|\treenode|=1$. $\mu_{0j}$ is a firing sequence of length
  at most $\maccovlen'(\guessfn(0j))$ that covers $\guessfn(0j)$ starting from $M_{0}$. The
  claim is clear by the definition of $\guessfn(0j)$.

  \emph{Induction step}: We want to prove that $\mu_{\treenode j}$
  can be fired at $M_{\treenode}$ and that $M_{\treenode j}$ satisfies
  the stated claims. We will prove these for an arbitrary place $p$.
  By induction hypothesis, either $M_{\treenode}(p)\ge
  M_{\treenode}'(p)$ or $M_{\treenode}(p)\ge
  \boundfn(|\treenode|-1,p)$.

  First, suppose that $M_{\treenode}(p)\ge M_{\treenode}'(p)$. Since
  $\mu_{\treenode j}$
  covers $\guessfn(\treenode j)$ starting from $M_{\treenode}'$,
  $M_{\treenode j}(p)\ge \guessfn(\treenode j)(p)$
  and there are no intermediate markings between $M_{\treenode}$ and
  $M_{\treenode j}$ where $p$ receives negative number of tokens. Also, since
  $M_{\treenode j}(p)\ge \guessfn(\treenode j)(p)$, either
  $M_{\treenode j}(p)\ge M_{\treenode j}'(p)$ or
  $M(\treenode j)(p)\ge \boundfn(|\treenode j|-1,p)$.

  Second, suppose that $M_{\treenode}(p)\ge
  \boundfn(|\treenode|-1,p)$. $|\mu_{\treenode j}|\le
  \maccovlen'(\guessfn(\treenode j))$ and $\guessfn(\treenode j)\le
  \boundfn(|\treenode j|-1)$ by definition. Hence
  $\maccovlen'(\guessfn(\treenode j))\le
  \maccovlen'(\boundfn(|\treenode j|-1))$ and $|\mu_{\treenode j}|\le
  \maccovlen'(\boundfn(|\treenode j|-1))$. By definition of
  $\boundfn(|\treenode|-1,p)$, we get $M_{\treenode}(p)\ge
  \wt\maccovlen'(\boundfn(|\treenode j|-1))+\boundfn(|\treenode j|-1,p)$.
  $\mu_{\treenode j}$
  will remove at most $\wt\maccovlen'(\boundfn(|\treenode j|-1))$ tokens from $p$ and hence, at least
  $\boundfn(|\treenode j|-1,p)$ tokens will be left in place $p$ at
  marking $M_{\treenode j}$.
  Therefore, $M_{\treenode j}(p)\ge \boundfn(|\treenode j|-1,p)$.

  This completes the induction and hence the claim.
  
  Now, we will prove  that each $M_{\treenode}$ satisfies
  $\nodecontent(\treenode)$. For
  each conjunct $\tau\ge c$ in $\nodecontent(\treenode)$, we will
  prove that $\sum_{p\in P}\linf_{\tau}(p)M_{\treenode}(p)\ge
  c$, where $\linf_{\tau}$ is the positive linear combination
  represented by $\tau$. If $c=0$, then the required
  result can be obtained by just observing that both
  $\linf_{\tau}(p)$ and $M_{\treenode}(p)$ are positive for all $p\in P$. So
  suppose that $c\ne 0$. Let $Q_{\tau}=\{p\in P\mid
  \linf_{\tau}(p)\ne 0\}$. We distinguish two cases:
  \begin{enumerate}
    \item For some $p\in Q_{\tau}$, $M_{\treenode}(p)\ge
      \boundfn(|\treenode|-1,p)$. In this
      case, $M_{\treenode}(p)\ge \boundfn(|\treenode|-1,p)\ge
      \frac{c}{\linf_{\tau}(p)}$. Hence,
      $\linf_{\tau}(p)M_{\treenode}(p)\ge c$.
    \item For all $p\in Q_{\tau}$,
      $M_{\treenode}(p)<\boundfn(|\treenode|-1,p)$. In this case,
      for all $p\in Q_{\tau}$, $M_{\treenode}(p)\ge M_{\treenode}'(p)$. Since
      $M_{\treenode}'$ satisfies $\nodecontent(\treenode)$, we have $\sum_{p\in
      Q_{\tau}}\linf_{\tau}(p)M_{\treenode}'(p)\ge c$. Therefore,
      $\sum_{p\in Q_{\tau}}\linf_{\tau}(p)M_{\treenode}(p)\ge c$.
  \end{enumerate}\qed
\end{proof}
To derive an upper bound for $\boundfn(i)$ to use in a nondeterministic
algorithm, let
$R=\max\{ratio(\tau \ge c) \mid \tau\ge
c\text{ is a subformula of }\kappa\}$,
$R'=R + \macmaxarcw + \macmaxarcw^{2} + \macmaxarcw^{3}$ and
$\wt'=\max\{\wt,2\}$. Recall that $\nestDepth-1$ is the nesting depth
of $\EF$ and note that boundedness and coverability can be expressed
with $\nestDepth\le 2$.
\begin{lemma}
  \label{lem:BoundOneip}
  For $i\ge 2$, $\boundfn(\nestDepth-i,p)\le (i+1)R'\wt\maccovlen'(\boundfn(\nestDepth-i+1))$.
\end{lemma}
  \begin{proof}
    By induction on $i$.

    \emph{Base case}: $i=2$
    \begin{align*}
      \boundfn(\nestDepth-2,p)&\le\text{max}\{R,\wt\maccovlen'(\boundfn(\nestDepth-1))+\boundfn(\nestDepth-1,p)\}\\
      &\le
      R+\wt\maccovlen'(\boundfn(\nestDepth-1))+\boundfn(\nestDepth-1,p)\\
      &\le R+\wt\maccovlen'(\boundfn(\nestDepth-1))+R\\
      &\le 2R+\wt\maccovlen'(\boundfn(\nestDepth-1))\\
      &\le 2R'\wt\maccovlen'(\boundfn(\nestDepth-1))
    \end{align*}

    \emph{Induction step}:
    \begin{align*}
      \boundfn(\nestDepth-i-1,p)&\le\text{max}\{R,\wt\maccovlen'(\boundfn(\nestDepth-i))+\boundfn(\nestDepth-i,p)\}\\
      &\le R+\wt\maccovlen'(\boundfn(\nestDepth-i))+(i+1)R'\wt\maccovlen'(\boundfn(\nestDepth-i+1))\\
      &\le R'\wt\maccovlen'(\boundfn(\nestDepth-i))+(i+1)R'\wt\maccovlen'(\boundfn(\nestDepth-i))\\
      &=(i+2)R'\wt\maccovlen'(\boundfn(\nestDepth-i))
    \end{align*}
  \end{proof}
\begin{lemma}
  \label{lem:BoundOnfofe}
  Let $q(i)=(2 \macnumplaces (\macvcnum' + 1)!)^{i}$. Then
  $\maccovlen'(\boundfn(\nestDepth-1))\le (2m\wt'R')^{q(1)}$ and also
  $\maccovlen'(\boundfn(\nestDepth-i))\le
  \prod_{j=\nestDepth-i}^{\nestDepth-1}\left(
  (\nestDepth-j+1)2 \macnumplaces\wt'^{8}R' \right)^{q(i+j+1-\nestDepth)}$.
\end{lemma}
\begin{proof}
  $\maccovlen'(\boundfn(\nestDepth-1))\le
  (2\macnumplaces \wt'R')^{q(1)}$ is by \lemref{lem:covLenUpBound}. Next result
  is by induction on $i$.

  \emph{Base case}: $i=2$. Since $\boundfn(\nestDepth-2,p)\le
  3R'\wt\maccovlen'(\boundfn(\nestDepth-1))$ and
  $\maccovlen'(\boundfn(\nestDepth-2))\le (2 \macnumplaces \wt r')^{q(1)}$ where
  $r'=\max\{\boundfn(\nestDepth-2,p)\mid p\in P\} + \wt +
  \wt^{2} + \wt^{3}$, we get
  \begin{align*}
    \maccovlen'(\boundfn(\nestDepth-2))&\le (2 \macnumplaces \wt
    ( 3R'\wt\maccovlen'(\boundfn(\nestDepth-1)) + \wt + \wt^{2} +
    \wt^{3}) )^{q(1)}\\
    &\le (3*2 \macnumplaces \wt'^{8} R')^{q(1)} ( 2 \macnumplaces \wt'
    R')^{q(2)}\\
  \end{align*}
  \emph{Induction step}: Since $\boundfn(\nestDepth-i-1,p)\le
  (i+2)R'\wt'\maccovlen'(\boundfn(\nestDepth-i))$, we have
  \begin{align*}
    \maccovlen'(\boundfn(\nestDepth-i-1))&\le (2 \macnumplaces \wt
    ((i+2)R'\wt'\maccovlen'(\boundfn(\nestDepth-i)) + \wt +
    \wt^{2} + \wt^{3}))^{q(1)}\\
    &\le \left(
    (i+2)2
    \macnumplaces\wt'^{8}R'\prod_{j=\nestDepth-i}^{\nestDepth-1}((\nestDepth-j+1)2
    \macnumplaces \wt'^{8}R')^{q(i+j+1-\nestDepth)}
    \right)^{q(1)}\\
    &=\left( (i+2)2 \macnumplaces \wt'^{8}R'
    \right)^{q(1)}\prod_{j=\nestDepth-i}^{\nestDepth-1}\left(
    (\nestDepth-j+1)2 \macnumplaces \wt'^{8}R' \right)^{q(i+1+j+1-\nestDepth)}\\
    &=\prod_{j=\nestDepth-i-1}^{\nestDepth-1}\left( (\nestDepth-j+1)2
    \macnumplaces \wt'^{8}R'
    \right)^{q(i+1+j+1-\nestDepth)}
  \end{align*}\qed
\end{proof}
\begin{theorem}
  \label{thm:kappaFormulaSatParaPspace}
  Given a Petri net with an initial marking and a $\kappa$ formula
  $\phi$, if the vertex cover number of the Petri net $\macvcnum$ and
  the maximum arc weight $\macmaxarcw$ are treated as parameters and
  the nesting depth $\nestDepth$ of $\EF$ modality in the formula is
  treated as a constant, then there is a \macparapsp{} algorithm that
  checks if the Petri net satisfies the given formula.
\end{theorem}
\begin{proof}
  First reduce $\phi$ to the form of $\gamma \land \EF( \kappa_{1})
  \land \cdots \land \EF( \kappa_{r})$, with $\gamma$ having only
  conjunctions of $\tau \ge \macicon$ formulas by nondeterministically
  choosing disjuncts from subformulas of $\phi$. By
  \lemref{lem:GuessFunctionPNPath}, it is enough for a
  nondeterministic algorithm to guess sequences $\sigma_{\treenode
  j}$, $\treenode j\in \treenodes$ of lengths at most
  $\maccovlen'(\boundfn(|\treenode j|-1))$ and verify that they
  satisfy the formula. Using bounds given by \lemref{lem:BoundOnfofe}
  and an argument similar to the one in the proof of
  \thmref{thm:covInParaPspace}, it can be shown that the space used
  is exponential in $\macvcnum'$ and polynomial in
  the size of the net and numeric constants in the formula. This gives
  the \macparapsp{} algorithm.\qed
\end{proof}
The space requirement of the above algorithm will have terms like
$\macnumplaces^{2\nestDepth}$ and hence it will not be \macparapsp{} if
$\nestDepth$ is treated as a parameter instead of a constant.

\subsection{Pumping sequences}
\label{sec:betaFormulas}
In order to check the truth of $\beta$ formulas, we adapt the concept
of \emph{disjointness sequence} introduced in \cite{SD2010} to our
notation. To make the presentation suitable for our setting, we use
terminology different from those used in \cite{SD2010}.
\begin{definition}[\cite{SD2010}]
  \label{def:pumpSeq}
  Let $\macubs \subseteq \macplaces$ be a non-empty subset of places.
  If $\macfso = \mactranso_{1} \cdots \mactranso_{\macfslidx}$ is a
  sequence of transitions and $\macplaceo$ is a place,
  $\maceff{\macfso}(\macplaceo)$ denotes the total effect of $\macfso$
  on $\macplaceo$: $\maceff{\macfso}(\macplaceo) =
  \sum_{\mactransidx=1}^{\macfslidx} \macpost(\macplaceo,
  \mactranso_{\mactransidx}) - \macpre(\macplaceo,
  \mactranso_{\mactransidx})$.  A firing sequence $\macfso$ enabled at
  an initial marking $\macmark_{0}: \macplaces \to \macnat$ is said to be a
  \macstandout{$\macubs$-pumping sequence} if $\macfso$ can be
  decomposed as
  $\macfso_{1}'\macpump{\macfso_{1}}\macfso_{2}' \macpump{\macfso_{2}}
  \cdots \macfso_{\macnumpart}' \macpump{\macfso_{\macnumpart}}$ such
  that
  \begin{enumerate}
    \item For each $\macplaceo\in \macplaces$,
      $\maceff{\macpump{\macfso_{1}}}(\macplaceo) \ge 0$ and for each
      $\macpartidx$ between $2$ and $\macnumpart$,
      $\maceff{\macpump{\macfso_{\macpartidx}}}(\macplaceo) < 0$
      implies there is a $\macpartidxt \le \macpartidx-1$ such that
      $\maceff{\macpump{\macfso_{\macpartidxt}}}(\macplaceo) > 0$ and
    \item $\macubs \subseteq \bigcup_{\macpartidx = 1}^{\macnumpart}
      \{\macplaceo \in \macplaces \mid
      \maceff{\macpump{\macfso_{\macpartidx}}}(\macplaceo) > 0\}$.
  \end{enumerate}
  The subsequences $\macpump{\macfso_{1}}, \dots,
  \macpump{\macfso_{\macnumpart}}$ are called pumping portions of the
  pumping sequence. They are underlined to distinguish them from
  non-pumping portions of the sequence.
\end{definition}
The following lemma from \cite{SD2010} establishes the connection
between model checking $\beta$ formulas and the existence of pumping
sequences.
\begin{lemma}[\cite{SD2010}]
  \label{lem:pumpSeqNecSuf}
  $\macnet,\macmark_{0}\models \{\tau_{1},\dots,\tau_{r}\}\macunbdd$ iff 
  there exists a $\macubs$-pumping sequence for some $\macubs
  \subseteq \macplaces$ such that for every $\macplaceidxt \in
  \{1,\cdots,r\}$, there is a $\macplaceo_{\macplaceidxt}\in \macubs$
  with $\maclinf_{\tau_{\macplaceidxt}}(\macplaceo_{\macplaceidxt})\ge 1$.
\end{lemma}
\begin{proof}
  ($\Leftarrow$) Suppose there is a $\macubs$-pumping sequence
  $\macfso$ as given in the lemma. Let
  $\macfso_{1}'\macpump{\macfso_{1}}
  \cdots \macfso_{\macnumpart}' \macpump{\macfso_{\macnumpart}}$ be the
  decomposition of $\macfso$ as in \defref{def:pumpSeq}. By repeating
  the subsequences
  $\macpump{\macfso_{1}},\dots,\macpump{\macfso_{\macnumpart}}$
  suitably many times (see \cite[Lemma 3.1]{SD2010}), we can ensure
  that for all $\macboundOnP \in \macnat$, there is a marking
  $\macmark$ reachable from $\macmark_{0}$ such that for all $\macplaceidxt \in
  \{1,\dots,r\}$, $\sum_{\macplaceo \in \macplaces}
  \linf_{\tau_{\macplaceidxt}}(\macplaceo) \macmark(\macplaceo) >
  \macboundOnP$.

  ($\Rightarrow$)  Suppose $\macnet,\macmark_{0}\models 
    \{\tau_{1},\dots,\tau_{r}\}\macunbdd$. By semantics, we get
    $\forall \macboundOnP\in \macnat$, there is a marking $\macmark$
    reachable from $\macmark_{0}$ such that for all $j\in
    \{1,\dots,r\}$ $\sum_{\macplaceo\in
    \macplaces}\linf_{\tau_{j}}(\macplaceo)\macmark(\macplaceo)>\macboundOnP$.
    Hence, we can conclude that for all $\macboundOnP\in \macnat$,
    there are places
    $\macplaceo_{1}^{\macboundOnP},\macplaceo_{2}^{\macboundOnP},\dots,\macplaceo_{r}^{\macboundOnP}$
    and $\macmark^{\macboundOnP}$ reachable from $\macmark_{0}$ such
    that $\macmark^{\macboundOnP}(p_{j}^{\macboundOnP})>\macboundOnP\land
    \linf_{\tau_{j}}(\macplaceo_{j}^{\macboundOnP})\ge 1$ for all $j\in
    \{1,\dots,r\}$. For each $\macboundOnP\in \macnat$, let
    $\macubs^{\macboundOnP}=\{\macplaceo_{1}^{\macboundOnP},\dots,\macplaceo_{r}^{\macboundOnP}\}$.
    Since the sequence $\macubs^{1}, \macubs^{2},\dots$ is infinite and there are
    only finitely many subsets of $\macplaces$, at least one subset of
    $\macplaces$
    occurs infinitely often in this sequence. Let $\macubs$ be this subset.
    We will now prove that there is a $\macubs$-pumping sequence using some
    results about coverability trees \cite[Section~4.6]{DR98}.

    Recall that in a coverability tree, markings $\macmark:\macplaces\to
    \macnat$ are extended to $\omega$-markings
    $\overline{\macmark}:\macplaces\to \macnat\cup\{\omega\}$, by
    mapping unbounded places to $\omega$. We first claim that there is
    some reachable $\omega$-marking $\overline{\macmark}$ in the
    coverability tree of $(\macnet,\macmark_{0})$ such that for all
    $\macplaceo\in \macubs$, $\overline{\macmark}(\macplaceo)=\omega$.
    Suppose not. Then, for every reachable $\omega$-marking
    $\overline{\macmark}$, there is some place $p\in \macubs$ such
    that $\overline{\macmark}(\macplaceo)<\omega$. Let $\macboundOnP$
    be the maximum of such bounds. Then, by \cite[Theorem 22]{DR98},
    for every marking $\macmark$ reachable from $\macmark_{0}$, there
    exists $\macplaceo\in \macubs$ such that $\macmark(\macplaceo)\le
    \macboundOnP$, a contradiction. Hence, there is a reachable
    $\omega$-marking $\overline{\macmark}$ in the coverability tree of
    $(\macnet,\macmark_{0})$ such that for all $\macplaceo\in
    \macubs$, $\overline{\macmark}(\macplaceo)=\omega$.  Now, the
    required $\macubs$-pumping sequence can be constructed (see
    \cite[Lemma 3.1]{SD2010} for details). \qed
\end{proof}

Model checking $\beta$ formulas thus reduces to detecting the presence
of certain $\macubs$-pumping sequences. The following definition
adapted from \cite{SD2010} is a generalization of
$\macplacesu$-enabled self-covering sequences.
\begin{definition}[\cite{SD2010}]
  \label{def:weakEnPumpSeq}
  Let $\macidplaces \subseteq \macplacesu \subseteq \macplaces$ be a
  subset of places that contains all independent places, $\macubst
  \subseteq \macplaces$ a possibly empty subset of places and $\macubs
  \subseteq \macplaces$ a non-empty subset of places. Let $\macmark:
  \macplaces \to \macint$ and $\macboundOnP \in \macnat \cup
  \{\omega\}$. A sequence of transitions is said to be a
  \macstandout{$\macubst$-neglecting weakly $\macmark, \macplacesu,
  \macboundOnP$-enabled $\macubs$-pumping sequence} if it can be
  decomposed as $\macfso_{1}'\macpump{\macfso_{1}}\macfso_{2}'
  \macpump{\macfso_{2}} \cdots \macfso_{\macnumpart}'
  \macpump{\macfso_{\macnumpart}}$ such that
  \begin{enumerate}
    \item For each $1 \le \macpartidx \le \macnumpart$, for each
      $\macplaceo \in \macplaces$, $\maceff{\macpump{
      \macfso_{\macpartidx}}} (\macplaceo) < 0$ implies (there is a
      $1\le \macpartidxt \le \macpartidx -1$ such that $\maceff{\macpump{
      \macfso_{\macpartidxt}}} (\macplaceo) > 0$ or $\macplaceo \in
      \macubst$).
    \item $\macubs \subseteq \bigcup_{\macpartidx = 1}^{\macnumpart}
      \{\macplaceo \in \macplaces \mid
      \maceff{\macpump{\macfso_{\macpartidx}}(\macplaceo)} > 0\}
      \setminus \macubst$.
    \item For any intermediate marking $\macmark'$ and any place
      $\macplaceo \in \macplacesu\setminus \macidplaces$, $\macmark'
      (\macplacesu) < \macboundOnP$.
    \item For any intermediate marking $\macmark'$ and any place
      $\macplaceo \in \macplacesu$, $\macmark' (\macplaceo) < 0$
      implies (there is a $\macpump{\macfso_{\macpartidxt}}$ occurring
      before $\macmark'$ such that $\maceff{ \macpump{
      \macfso_{\macpartidxt}}} (\macplaceo) > 0$ or $\macplaceo \in
      \macubst$).
  \end{enumerate}
\end{definition}
Intuitively, a $\macubst$-neglecting weakly $\macmark, \macplacesu,
\macboundOnP$-enabled $\macubs$-pumping sequence maintains the number
of tokens between $0$ and $\macboundOnP$ in all places in
$\macplacesu$ while in other places, it can become less than $0$ or
more than $\macboundOnP$. If a place $\macplaceo \in \macplacesu$ has
already been pumped up by some pumping portion
$\macpump{\macfso_{\macpartidxt}}$, $\macplaceo$ may have negative
number of tokens in intermediate markings that occur after $\macpump{
\macfso_{\macpartidxt}}$. The following lemma implies that for
detecting the presence of pumping sequences, it is enough to detect
certain weakly enabled pumping sequences.

\begin{lemma}[\cite{SD2010}]
  \label{lem:WeakEnSeqSuf}
  Let $\macubs \subseteq \macplaces$ be a non-empty subset of places
  and $\macmark_{0}: \macplaces \to \macnat$ be the initial marking.
  Any $\macubs$-pumping sequence enabled at $\macmark_{0}$ is a
  $\emptyset$-neglecting weakly $\macmark_{0}, \macplaces,
  \omega$-enabled $\macubs$-pumping sequence. Suppose that $\macfso =
  \macfso_{1}'\macpump{\macfso_{1}}\macfso_{2}' \macpump{\macfso_{2}}
  \cdots \macfso_{\macnumpart}' \macpump{\macfso_{\macnumpart}}$ is a
  $\emptyset$-neglecting weakly $\macmark_{0}, \macplaces,
  \omega$-enabled $\macubs$-pumping sequence. Then, there are
  $\macnumrep_{1}, \macnumrep_{2}, \dots, \macnumrep_{\macnumpart} \in
  \macnat$ such that
  $\macfso_{1}'\macpump{\macfso_{1}}^{\macnumrep_{1}} \macfso_{2}'
  \macpump{\macfso_{2}}^{\macnumrep_{2}} \cdots \macfso_{\macnumpart}'
  \macpump{\macfso_{\macnumpart}}^{\macnumrep_{\macnumpart}}$ is a
  $\macubs$-pumping sequence enabled at $\macmark_{0}$.
\end{lemma}
\begin{proof}
  The first part follows from definitions. For the second part, we
  define $\macnumrep_{\macnumpart}, \dots, \macnumrep_{1}$ in that
  order as follows:
  \begin{itemize}
    \item $\macnumrep_{\macnumpart} = 1$.  \item Suppose $1\le
      \macpartidx < \macnumpart$ and $\macnumrep_{\macpartidx+1},
      \dots, \macnumrep_{\macnumpart}$ have already been defined.
      Define $\macnumrep_{\macpartidx}$ to be $(\macnumpart -
      \macpartidx) (|\macfso|-1)\macmaxarcw + \sum_{\macpartidxt =
      \macpartidx+1}^{\macnumpart} (|\macfso|-1) \macmaxarcw
      \macnumrep_{\macpartidxt}$.
  \end{itemize}
  We will prove that $\macfso' =
  \macfso_{1}'\macpump{\macfso_{1}}^{\macnumrep_{1}} \macfso_{2}'
  \macpump{\macfso_{2}}^{\macnumrep_{2}} \cdots \macfso_{\macnumpart}'
  \macpump{\macfso_{\macnumpart}}^{\macnumrep_{\macnumpart}}$
  satisfies all conditions of \defref{def:pumpSeq} and that it is
  enabled at $\macmark_{0}$. Condition 2 follows by the fact that
  $\macfso$ satisfies condition 2 of \defref{def:weakEnPumpSeq} and
  that $\macubst = \emptyset$.  Condition 1 of \defref{def:pumpSeq}
  follows by the fact that $\macfso$ satisfies condition 1 of
  \defref{def:weakEnPumpSeq} and that $\macubst = \emptyset$. For
  proving that $\macfso'$ is enabled at $\macmark_{0}$, we will prove
  the following claim by induction on $\macpartidx$: for any
  intermediate marking $\macmark'$ occurring when firing $\macfso_{1}'
  \macpump{\macfso_{1}}^{\macnumrep_{1}} \cdots \macfso_{\macpartidx}'
  \macpump{ \macfso_{\macpartidx}}^{\macnumrep_{\macpartidx}}$ from
  $\macmark_{0}$ and any $\macplaceo \in \macplaces$, $\macmark'
  (\macplaceo) \ge 0$; and for any intermediate marking $\macmark''$
  occurring while firing $\macfso'$ from $\macmark_{0}$ and any
  $\macplaceo' \in \bigcup_{\macpartidxt = 1}^{\macpartidx}
  \{\macplaceo \in \macplaces \mid \maceff{\macpump{
  \macfso_{\macpartidxt}}} (\macplaceo) > 0\}$,
  $\macmark''(\macplaceo) \ge 0$.

  Base case $\macpartidx = 1$: Since $\macubst = \emptyset$ and
  $\macfso$ satisfies condition 4 of \defref{def:weakEnPumpSeq}, for
  any intermediate marking $\macmark'$ occurring when firing
  $\macfso_{1}' \macpump{\macfso_{1}}$ from $\macmark_{0}$ and any
  place $\macplaceo \in \macplaces$, $\macmark' (\macplaceo) \ge 0$.
  Since $\macfso$ satisfies condition 1 of \defref{def:weakEnPumpSeq}
  and $\macubst = \emptyset$, $\maceff{ \macpump{ \macfso_{1}}}(
  \macplaceo) \ge 0$ for any place $\macplaceo \in \macplaces$. Hence,
  for any intermediate marking $\macmark'$ occurring when firing
  $\macfso_{1}' \macpump{\macfso_{1}}^{\macnumrep_{1}}$ from
  $\macmark_{0}$ and any place $\macplaceo \in \macplaces$, $\macmark'
  (\macplaceo) \ge 0$. Since $|\macfso_{2}'\cdots
  \macfso_{\macnumpart}'| \le (\macnumpart - 1)(|\macfso|-1)$ and
  $|\macpump{ \macfso_{2}}^{\macnumrep_{2}}\cdots \macpump{
  \macfso_{\macnumpart}}^{\macnumrep_{\macnumpart}}| \le
  \sum_{\macpartidxt = 2}^{\macnumpart}
  (|\macfso|-1)\macnumrep_{\macpartidxt}$, $\macfso_{2}'
  \macpump{ \macfso_{2}}^{\macnumrep_{2}} \cdots
  \macfso_{\macnumpart}' \macpump{
  \macfso_{\macnumpart}}^{\macnumrep_{\macnumpart}}$ can decrease at
  most $(\macnumpart - 1) (|\macfso|-1)\macmaxarcw +
  \sum_{\macpartidxt = 2}^{\macnumpart} (|\macfso|-1) \macmaxarcw
  \macnumrep_{\macpartidxt}$ tokens from any place. If $\macmark_{0}
  \macStep{\macfso_{1} \macpump{ \macfso_{1}}^{\macnumrep_1}}
  \macmark_{1}$ and $\maceff{ \macpump{ \macfso_{1}}} (\macplaceo) >
  0$ for any place $\macplaceo$, then $\macmark_{1}(\macplaceo) \ge
  (\macnumpart - 1) (|\macfso|-1)\macmaxarcw +
  \sum_{\macpartidxt = 2}^{\macnumpart} (|\macfso|-1) \macmaxarcw
  \macnumrep_{\macpartidxt}$. Hence, the second part of the claim follows.

  Induction step: Assume that $\macmark_{0} \macStep{
  \macfso_{1}' \macpump{ \macfso_{1}}^{\macnumrep_{1}} \cdots
  \macfso_{\macpartidx}' \macpump{
  \macfso_{\macpartidx}}^{\macnumrep_{\macpartidx}}}
  \macmark_{\macpartidx}$. Suppose for some place $\macplaceo'$ and
  some intermediate marking $\macmark'$ that occurs while firing
  $\macfso_{\macpartidx + 1} \macpump{
  \macfso_{\macpartidx+1}}$ from $\macmark_{\macpartidx}$, $\macmark'
  (\macplaceo) < 0$. By induction hypothesis, $\macplaceo' \notin
  \bigcup_{\macpartidxt = 1}^{\macpartidx} \{\macplaceo \in \macplaces
  \mid \maceff{\macpump{ \macfso_{\macpartidxt}}} (\macplaceo) > 0\}$,
  which contradicts the fact that $\macfso$ satisfies conditions 1 and
  4 of \defref{def:weakEnPumpSeq}. Also from condition 1 of
  \defref{def:weakEnPumpSeq}, $\maceff{ \macpump{
  \macfso_{\macpartidx+1}}} (\macplaceo) \ge 0$ for any $\macplaceo
  \notin \bigcup_{\macpartidxt = 1}^{\macpartidx} \{\macplaceo \in \macplaces
  \mid \maceff{\macpump{ \macfso_{\macpartidxt}}} (\macplaceo) > 0\}$.
  Hence, for all $\macplaceo \in \macplaces$ and any intermediate
  marking $\macmark'$ that occurs while firing
  $\macfso_{\macpartidx+1}' \macpump{
  \macfso_{\macpartidx+1}}^{\macnumrep_{\macpartidx+1}}$ from
  $\macmark_{\macpartidx}$, $\macmark' (\macplaceo) \ge 0$. Suppose
  $\macpartidx+2 \le \macnumpart$. Since $|\macfso_{\macpartidx+2}'\cdots
  \macfso_{\macnumpart}'| \le (\macnumpart - \macpartidx -1)(|\macfso|-1)$ and
  $|\macpump{ \macfso_{\macpartidx+2}}^{\macnumrep_{\macpartidx+2}}\cdots \macpump{
  \macfso_{\macnumpart}}^{\macnumrep_{\macnumpart}}| \le
  \sum_{\macpartidxt = \macpartidx+2}^{\macnumpart}
  (|\macfso|-1)\macnumrep_{\macpartidxt}$, $\macfso_{\macpartidx+2}'
  \macpump{ \macfso_{\macpartidx+2}}^{\macnumrep_{\macpartidx+2}} \cdots
  \macfso_{\macnumpart}' \macpump{
  \macfso_{\macnumpart}}^{\macnumrep_{\macnumpart}}$ can decrease at
  most $(\macnumpart - \macpartidx - 1) (|\macfso|-1)\macmaxarcw +
  \sum_{\macpartidxt = \macpartidx + 2}^{\macnumpart} (|\macfso|-1) \macmaxarcw
  \macnumrep_{\macpartidxt}$ tokens from any place. If
  $\macmark_{\macpartidx} \macStep{\macfso_{\macpartidx+1}'
  \macpump{ \macfso_{\macpartidx+1}}^{\macnumrep_{\macpartidx+1}}}
  \macmark_{\macpartidx+1}$ and $\maceff{
  \macpump{ \macfso_{\macpartidx+1}}} (\macplaceo) >
  0$ for any place $\macplaceo$, then $\macmark_{\macpartidx+1}(\macplaceo) \ge
  (\macnumpart - \macpartidx - 1) (|\macfso|-1)\macmaxarcw +
  \sum_{\macpartidxt = \macpartidx + 2}^{\macnumpart} (|\macfso|-1) \macmaxarcw
  \macnumrep_{\macpartidxt}$. Hence, second part of the claim follows.
  \qed
\end{proof}

As is done in \secref{sec:boundedness}, we will bound the length of
weakly enabled pumping sequences by induction on $|\macplacesu|$. The
following two lemmas are helpful in manipulating weakly enabled
pumping sequences.

\begin{lemma}[\cite{SD2010}]
  \label{lem:weakPupmSeqRep}
  Suppose $\macfso = \macfso_{1}' \macpump{\macfso_{1}} \macfso_{2}'
  \cdots \macfso_{\macnumpart}' \macpump{ \macfso_{\macnumpart}}$ is a
  $\macubst$-neglecting $\macmark, \macplacesu, \omega$-enabled
  $\macubs$-pumping sequence. Then the sequence $\macfso' =
  \macfso_{1}' \macfso_{1}^{\macnumrep_{1}} \macpump{\macfso_{1}}
  \macfso_{1}^{\macnumrep_{1}'} \macfso_{2}' \cdots
  \macfso_{\macnumpart}'
  \macfso_{\macnumpart}^{\macnumrep_{\macnumpart}} \macpump{
  \macfso_{\macnumpart}}$ is also a $\macubst$-neglecting $\macmark,
  \macplacesu, \omega$-enabled $\macubs$-pumping sequence for any
  $\macnumrep_{1}, \macnumrep_{1}', \dots, \macnumrep_{\macnumpart}
  \in \macnat$ ($\macfso_{\macpartidx}$ is same as $\macpump{
  \macfso_{\macpartidx}}$, except that $\macfso_{\macpartidx}$ is not
  considered a pumping portion while $\macpump{
  \macfso_{\macpartidx}}$ is considered a pumping portion).
\end{lemma}
\begin{proof}
  We will prove that the new sequence satisfies all the conditions of
  \defref{def:weakEnPumpSeq}. Conditions 1 and 2 are satisfied since
  the set of pumping portions of the new sequence is equal to that
  of the old one and occurs in the same order. Condition 3 is
  trivially satisfied since in this case, $\macboundOnP = \omega$.
  Suppose for some intermediate marking $\macmark'$ and some place
  $\macplaceo \in \macplacesu$, $\macmark' (\macplaceo) < 0$. Let
  $\macpartidxt$ be the maximum number such that $\macpump{
  \macfso_{\macpartidxt}}$ occurs before $\macmark'$. Suppose
  $\macmark \macstep{ \macfso_{1}'\macfso_{1}^{\macnumrep_{1}} \macpump{\macfso_{1}}
  \macfso_{1}^{\macnumrep_{1}'} \macfso_{2}' \cdots
  \macfso_{\macpartidxt}'
  \macfso_{\macpartidxt}^{\macnumrep_{\macpartidxt}} \macpump{
  \macfso_{\macpartidxt}}
  }{} \macmark''$
  and $\macmark'' \macstep{\macfsh}{} \macmark'$. If $\macplaceo \in
  \macubst$ or $\macplaceo \in \bigcup_{\macpartidxt' =
  1}^{\macpartidxt}\{\macplaceo' \in \macplaces \mid \maceff{
  \macpump{ \macfso_{\macpartidxt'}}}(\macplaceo') > 0\}$, there is
  nothing else to prove. Otherwise, $\maceff{
  \macpump{ \macfso_{\macpartidxt'}}}(\macplaceo) = 0$ for every
  $\macpartidxt'$ between $1$ and $\macpartidxt$. This implies that if $\macmark \macstep{
  \macfso_{1}' \macpump{\macfso_{1}} \macfso_{2}' \cdots
  \macfso_{\macpartidxt}' \macpump{ \macfso_{\macpartidxt}} }{}
  \macmark_{2}$ and $\macmark_{2} \macstep{\macfsh}{}
  \macmark_{3}$, then $\macmark_{3} (\macplaceo) < 0$,
  contradicting the fact that $\macfso$ satisfies condition 4 of
  \defref{def:weakEnPumpSeq}.\qed
\end{proof}
\begin{lemma}
  \label{lem:weakPumpSeqCombine}
  Suppose $\macfso = \macfso_{1}' \macpump{\macfso_{1}} \cdots
  \macfso_{\macnumpart}' \macpump{\macfso_{\macnumpart}}$ is a
  $\macubst$-neglecting weakly $\macmark, \macplacesu, \omega$-enabled
  $\macubs_{1}$-pumping sequence and $\macfst = \macfst_{1}'
  \macpump{\macfst_{1}} \cdots \macfst_{\macnumpart'}'
  \macpump{ \macfst_{\macnumpart'}}$ is a $\macubst_{1}$-neglecting
  weakly $\macmark_{1}, \macplacesu, \omega$-enabled
  $\macubs_{2}$-pumping sequence. If $\macubst_{1} = \macubst \cup
  \{\macplaceo \in \macplaces \mid \maceff{ \macpump{
  \macfso_{\macpartidx}}} (\macplaceo) > 0, 1\le \macpartidx \le
  \macnumpart\}$, $\macmark \macstep{\macfso}{} \macmark_{2}$ and for
  all $\macplaceo \in \macplacesu \setminus \macubst_{1}$,
  $\macmark_{2}(\macplaceo) = \macmark_{1}(\macplaceo)$, then $\macfso
  \macfst = \macfso_{1}' \macpump{\macfso_{1}} \cdots
  \macfso_{\macnumpart}' \macpump{\macfso_{\macnumpart}} \macfst_{1}'
  \macpump{\macfst_{1}} \cdots \macfst_{\macnumpart'}' \macpump{
  \macfst_{\macnumpart'}}$ is  a $\macubst$-neglecting weakly
  $\macmark,\macplacesu, \omega$-enabled $(\macubs_{1} \cup
  \macubs_{2})$-pumping sequence.
\end{lemma}
\begin{proof}
  We will prove that the combined sequence satisfies all conditions of
  \defref{def:weakEnPumpSeq}.
  \begin{enumerate}
    \item This follows since $\macfso$ and $\macfst$ individually
      satisfy condition 1 of \defref{def:weakEnPumpSeq} and
      $\macubst_{1} = \macubst \cup \{\macplaceo \in \macplaces \mid
      \maceff{ \macpump{ \macfso_{\macpartidx}}} (\macplaceo) > 0,
      1\le \macpartidx \le \macnumpart\}$.
    \item This follows from the fact that $\macubs_{1}$ and
      $\macubs_{2}$ individually satisfy condition 2 of
      \defref{def:weakEnPumpSeq}.
    \item This is trivially satisfied since in this case,
      $\macboundOnP = \omega$.
    \item Suppose $\macmark'$ is some intermediate marking that occurs
      while firing $\macfst$ from $\macmark_{2}$ with $\macmark'
      (\macplaceo) < 0$ for some $\macplaceo \in \macplacesu$. If
      $\macplaceo \in \macubst_{1}$ or there is some $\macpump{
      \macfst_{\macpartidx'}}$ occurring before $\macmark'$ such that
      $\maceff{\macpump{ \macfst_{\macpartidx'}}} (\macplaceo) > 0$,
      there is nothing more to prove. Otherwise, the fact that
      $\macplaceo \in \macplacesu \setminus \macubst_{1}$ and
      $\macmark_{2} (\macplaceo) = \macmark_{1} (\macplaceo)$
      contradicts the fact that $\macfst$ is a
      $\macubst_{1}$-neglecting weakly $\macmark, \macplacesu,
      \omega$-enabled $\macubs_{2}$-pumping sequence, that should have
      satisfied condition 4 of \defref{def:weakEnPumpSeq}.\qed
  \end{enumerate}
\end{proof}

Now, we will generelize $\macslencov$ and $\macscovlen$ to weakly
enabled pumping sequences so that we can calculate bounds on their
lengths by induction on $|\macplacesu|$. 
\begin{definition}
  \label{def:placeCaringZone}
  Let $\macplacesu, \macubs, \macubst \subseteq \macplaces$ be subsets
  of places such that $\macidplaces \subseteq \macplacesu$ and $\macubs$
  is non-empty. Suppose $\macfso = \macfso_{1}' \macpump{ \macfso_{1}}
  \cdots \macfso_{\macnumpart}' \macpump{ \macfso_{\macnumpart}}$ is a
  $\macubst$-neglecting weakly $\macmark, \macplacesu, \omega$-enabled
  $\macubs$-pumping sequence for some $\macmark: \macplaces \to
  \macint$.  For some independent place $\macplaceo \in \macidplaces$,
  if there is a $\macpartidxt$ such that $\maceff{ \macpump{
  \macfso_{\macpartidxt}}} > 0$, we do not care if $\macplaceo$ has
  negative number of tokens in some intermediate marking that occurs
  after $\macpump{ \macfso_{\macpartidxt}}$, even if $\macplaceo
  \notin \macubst$. For each $\macplaceo \in \macidplaces \setminus
  \macubst$, let $\macpartidxt[\macplaceo]$ be the minimum number such
  that $\maceff{ \macpump{ \macfso_{\macpartidxt[\macplaceo]}}}
  (\macplaceo) > 0$. If $\macmark \macstep{\macfso_{1}' \macpump{
  \macfso_{1}} \cdots \macfso_{\macpartidxt[\macplaceo]}' \macpump{
  \macfso_{\macpartidxt[\macplaceo]}}}{} \macmark_{1}$, then the set
  of all intermediate markings occurring between $\macmark$ and
  $\macmark_{1}$ (including $\macmark$ and $\macmark_{1}$) is called
  the \macstandout{caring zone of $\macplaceo$}. If there is no
  $\macpump{ \macfso_{\macpartidxt}}$ such that $\maceff{ \macpump{
  \macfso_{\macpartidxt}}} ( \macplaceo) > 0$, then the caring zone of
  $\macplaceo$ is the set of all intermediate markings.
\end{definition}

\begin{definition}
  \label{def:pumLenBound}
  Let $\macmaxinit' \in \macnat$ be some fixed number. For $\macidnidx
  \in \macnat$, $\macplacesu, \macubs, \macubst \subseteq \macplaces$
  with $\macidplaces \subseteq \macplacesu$ and $\macubs$ non-empty
  and a function $\macmark: \macplaces \to \macint$,
  $\macpumlen(\macplacesu, \macidnidx, \macmark, \macubs, \macubst)$
  is the length of the shortest $\macubst$-neglecting weakly
  $\macmark, \macplacesu, \omega$-enabled $\macubs$-pumping sequence
  from $\macmark$ if there is a $\macubst$-neglecting weakly
  $\macmark, \macplacesu, \omega$-enabled $\macubs$-pumping sequence
  from $\macmark$ in which every independent place $\macplaceo \in
  \macidplaces \setminus \macubst$ has at most $\macmaxinit' +
  \macidnidx \macmaxarcw$ tokens in all intermediate markings
  belonging to the caring zone of $\macplaceo$. Let
  $\macpumlen(\macplacesu, \macidnidx, \macmark, \macubs, \macubst)$
  be $0$ if there is no such sequence. Let
  $\macpumseqlen(\macplaceidx, \macidnidx) = \max
  \{\macpumlen(\macplacesu, \macidnidx, \macmark, \macubs, \macubst)
  \mid \macidplaces \subseteq \macplacesu \subseteq \macplaces,
  |\macplacesu \setminus \macidplaces| = \macplaceidx,
  \macmark:\macplaces \to \macint, \macubs,\macubst \subseteq
  \macplaces, \macubs \ne \emptyset\}$.
\end{definition}

\begin{lemma}
  \label{lem:shortPumSeq}
  Let $\macplacesu, \macubs, \macubst \subseteq \macplaces$ be subsets
  of places such that $\macidplaces \subseteq \macplacesu$ and
  $\macubs$ is non-empty and let $\macmaxinit' \in \macnat$. Let
  $\macnumtok \in \macnat$. Suppose there is a $\macubst$-neglecting
  weakly $\macmark, \macplacesu, \macnumtok$-enabled $\macubs$-pumping
  sequence $\macfso = \macfso_{1}' \macpump{\macfso_{1}} \cdots
  \macfso_{\macnumpart}' \macpump{\macfso_{\macnumpart}}$ for some
  $\macmark: \macplaces \to \macint$ such that every place $\macplaceo
  \in \macidplaces \setminus \macubst$ has at most $\macmaxinit'$
  tokens in all intermediate markings belonging to the caring zone of
  $\macplaceo$. Then, there is a $\macubst$-neglecting weakly
  $\macmark, \macplacesu, \omega$-enabled $\macubs$-pumping sequence
  of length at most $8 \macnumpart \macvcnum' (2
  \macnumtok)^{\macicon'\macvcnum'^{3}} (\macmaxinit'
  \macmaxarcw)^{\macicon' \macnumplaces^{4}}$ for some constant
  $\macicon'$.
\end{lemma}
\begin{proof}
  By induction on $\macnumpart$.

  Base case $\macnumpart = 1$: In this case, $\macfso =
  \macfso_{1}' \macpump{\macfso_{1}}$. All intermediate markings
  occurring as a result of firing $\macfso$ from $\macmark$ belong to
  the caring zone of each place $\macplaceo \in \macidplaces \setminus
  \macubst$. If any two intermediate
  markings occurring when $\macfso_{1}'$ is fired from $\macmark$
  agree on all places in $\macplacesu \setminus \macubst$, then the
  subsequence between them can be removed. Hence, we can assume
  without loss of generality that $|\macfso_{1}'| \le
  \macmaxinit'^{\macnumplaces} \macnumtok^{\macvcnum'}$.

  As in Rackoff's proof of Lemma 4.5 in \cite{RCK78}, remove
  $\macplacesu \setminus \macubst$-loops from $\macpump{\macfso_{1}}$
  carefully until what remains behind is a sequence $\macfso_{1}''$ of
  length at most
  $(\macmaxinit'^{\macnumplaces}\macnumtok^{\macvcnum'}+1)^{2}$. Let
  $\macpupcv\in \macnat^{|\macspplaces \setminus \macubst|}$ be the
  vector containing a $1$ in each coordinate corresponding to a
  special place in $\macspplaces \setminus \macubst$ whose number of
  tokens is increased by $\macpump{\macfso_{1}}$ and $0$ in all other
  coordinates. If $\macfirseqt$ is a $\macplacesu \setminus
  \macubst$-loop, its \macstandout{loop value} is the vector in
  $\macint^{|\macspplaces \setminus \macubst|}$, which contains in
  each coordinate the total effect of $\macfirseqt$ on the
  corresponding special place in $\macspplaces \setminus \macubst$.
  Let $\maclvs\subseteq \macint^{|\macspplaces \setminus \macubst|}$
  be the set of loop values that were removed from
  $\macpump{\macfso_{1}}$. Let $\maclvm$ be the matrix with
  $|\macspplaces \setminus \macubst|$ rows, whose columns are the
  members of $\maclvs$. For any sequence $\macfirseqt$, let
  $\macef(\macfirseqt)$ be the vector in $\macint^{|\macspplaces
  \setminus \macubst|}$, which contains in each coordinate the total
  effect of $\macfirseqt$ on the corresponding special place in
  $\macspplaces \setminus \macubst$. By definition,
  $\macef(\macpump{\macfso_{1}})\ge \macpupcv$. The effect of
  $\macpump{\macfso_{1}}$ can be split into the effect of
  $\macfso_{1}''$ and the effect of $\macplacesu\setminus
  \macubst$-loops that were removed from $\macpump{\macfso_{1}}$. If
  $\macsolv(\maclvidx)$ is the number of $\macplacesu\setminus
  \macubst$-loops removed from $\macpump{\macfso_{1}}$ whose loop
  value is equal to the $\maclvidx$\textsuperscript{th} column of
  $\maclvm$, then we have $\maclvm\macsolv\ge
  \macpupcv-\macef(\macfso_{1}'')$.

  A loop value is just the effect of at most
  $\macnumtok^{\macvcnum'}\macmaxinit'^{\macnumplaces}$ transitions,
  and hence each entry of $\maclvm$ is of absolute value at most
  $\macnumtok^{\macvcnum'}\macmaxinit'^{\macnumplaces}\macmaxarcw$.
  The matrix $\maclvm$ has therefore at most
  $(2\macnumtok^{\macvcnum'}\macmaxinit'^{\macnumplaces}\macmaxarcw+1)^{\macvcnum'}$
  columns. Each entry of $\macpupcv-\macef(\macfso_{1}'')$ is of
  absolute value at most
  $\macmaxarcw(\macnumtok^{\macvcnum'}\macmaxinit'^{\macnumplaces}+1)^{2}+1$.
  Letting $d_{1}=\macvcnum'$ and $d=\max\{
  (2\macnumtok^{\macvcnum'}\macmaxinit'^{\macnumplaces}\macmaxarcw+1)^{\macvcnum'},
  \macnumtok^{\macvcnum'}\macmaxinit'^{\macnumplaces}\macmaxarcw,
  \macmaxarcw(\macnumtok^{\macvcnum'}\macmaxinit'^{\macnumplaces}+1)^{2}+1\}\le
  (2\macnumtok)^{3\macvcnum'}(\macmaxinit'\macmaxarcw)^{3\macnumplaces^{2}}$,
  we can apply Lemma 4.4 of \cite{RCK78}. The result is that there is
  a vector $\macssolv\in \macnat^{|\maclvs|}$ such that the sum of
  entries of $\macssolv$ is equal to $l_{1}\le d(
  (2\macnumtok)^{3\macvcnum'}
  (\macmaxinit'\macmaxarcw)^{3\macnumplaces^{2}})^{\macicon\macvcnum'}$
  for some constant $\macicon$. Let $\macicon'$ be a constant such
  that $l_{1}\le \macvcnum' (2\macnumtok)^{\macicon'\macvcnum'^{2}}
  (\macmaxinit'\macmaxarcw)^{\macicon'\macnumplaces^{3}}$.

  Now, we will put back $l_{1}$ $\macplacesu \setminus \macubst$-loops
  back to $\macfso_{1}''$, which was of length at most
  $(\macnumtok^{\macvcnum'}\macmaxinit'^{\macnumplaces}+1)^{2}$. Since
  the length of each $\macplacesu \setminus \macubst$-loop is at most
  $\macnumtok^{\macvcnum'}\macmaxinit'^{\macnumplaces}$, the total
  length of the newly constructed pumping portion is at most
  $(\macnumtok^{\macvcnum'}\macmaxinit'^{\macnumplaces}+1)^{2}+\macvcnum'
  (2\macnumtok)^{\macicon'\macvcnum'^{3}}(\macmaxinit'\macmaxarcw)^{\macicon'\macnumplaces^{4}}$.
  Together with $\macfirseqo_{1}$, whose length is at most
  $\macnumtok^{\macvcnum'}\macmaxinit'^{\macnumplaces}$, we get a
  $\macubst$-neglecting weakly $\macmark, \macplacesu, \omega$-enabled
  $\macubs$-pumping sequence of length at most
  $2(\macnumtok^{\macvcnum'}\macmaxinit'^{\macnumplaces}+1)^{2}+\macvcnum'
  (2\macnumtok)^{\macicon'\macvcnum'^{3}}
  (\macmaxinit'\macmaxarcw)^{\macicon'\macnumplaces^{4}} \le
  8\macvcnum' (2\macnumtok)^{\macicon'\macvcnum'^{3}}
  (\macmaxinit'\macmaxarcw)^{\macicon'\macnumplaces^{4}}$.
  
  Induction step: Suppose $\macfso = \macfso_{1}' \macpump{
  \macfso_{1}} \cdots \macfso_{\macnumpart+1}' \macpump{
  \macfso_{\macnumpart+1}}$. Let $\macubs_{1} = \{\macplaceo \in
  \macplaces \mid \maceff{ \macpump{\macfso_{1}}}
  (\macplaceo) > 0 \}$. The sequence $\macfso_{1}' \macpump{
  \macfso_{1}}$ is a $\macubst$-neglecting weakly $\macmark,
  \macplacesu, \omega$-enabled $\macubs_{1}$-pumping sequence. Let
  $\macmark \macstep{\macfso_{1}' \macpump{ \macfso_{1}}}{}
  \macmark_{1}$. As is done in the base case, we can replace
  $\macfso_{1}' \macpump{ \macfso_{1}}$ by another
  $\macubst$-neglecting weakly $\macmark, \macplacesu, \omega$-enabled
  $\macubs_{1}$-pumping sequence $\macfso'$ of length at most
  $8\macvcnum' (2\macnumtok)^{\macicon'\macvcnum'^{3}}
  (\macmaxinit'\macmaxarcw)^{\macicon'\macnumplaces^{4}}$ ending at
  some marking $\macmark_{2}$ such that for all $\macplaceo \in
  \macplacesu \setminus \macubst$, $\macmark_{2} (\macplaceo) =
  \macmark_{1} (\macplaceo)$ (this is because we only remove
  $\macplacesu \setminus \macubst$ loops from $\macfso_{1}'
  \macpump{\macfso_{1}}$ to obtain the shorter sequence $\macfso'$).

  The sequence $\macfso_{2}' \macpump{ \macfso_{2}} \cdots
  \macfso_{\macnumpart+1}' \macpump{ \macfso_{\macnumpart+1}}$ is a
  $(\macubst \cup \macubs_{1})$-neglecting weakly $\macmark_{1},
  \macplacesu, \omega$-enabled $(\macubs \setminus
  \macubs_{1})$-pumping sequence. By induction hypothesis, there is
  another $(\macubst \cup \macubs_{1})$-neglecting weakly $\macmark_{1},
  \macplacesu, \omega$-enabled $(\macubs \setminus
  \macubs_{1})$-pumping sequence $\macfso''$ of length at most
  $8\macvcnum' \macnumpart(2\macnumtok)^{\macicon'\macvcnum'^{3}}
  (\macmaxinit'\macmaxarcw)^{\macicon'\macnumplaces^{4}}$.
  \Lemref{lem:weakPumpSeqCombine} implies that $\macfso' \macfso''$ is
  a $\macubst$-neglecting weakly $\macmark, \macplacesu,
  \omega$-enabled $(\macubs \setminus \macubs_{1}) \cup
  \macubs_{1}$-pumping sequence. The length of $\macfso' \macfso''$ is
  at most $8\macvcnum' (\macnumpart + 1)
  (2\macnumtok)^{\macicon'\macvcnum'^{3}}
  (\macmaxinit'\macmaxarcw)^{\macicon'\macnumplaces^{4}}$. \qed
\end{proof}

Using the technical lemmas proved above, we will now obtain a
recurrence relation for $\macpumseqlen$.
\begin{lemma}
  \label{lem:weakPumSeqRecRelBase}
  $\macpumseqlen(0, \macidnidx) \le 8 \macnumplaces
  \macvcnum' (2 (\macmaxinit' + \macidnidx \macmaxarcw) \macmaxarcw)^{\macicon'
  \macnumplaces^{4}}$.
\end{lemma}
\begin{proof}
  By \lemref{lem:shortPumSeq} after setting $\macnumtok = 1$ and
  substituting $\macmaxinit'$ by $\macmaxinit' + \macidnidx
  \macmaxarcw$.
\end{proof}
\begin{lemma}
  \label{lem:weakPumSeqRecRelInd}
  $\macpumseqlen(\macplaceidx+1, \macidnidx) \le 10
  \macnumplaces \macvcnum' (2 \macmaxarcw \macpumseqlen( \macplaceidx,
  \macidnidx+1))^{\macicon' \macvcnum'^{3}} ( (\macmaxinit'+
  \macidnidx \macmaxarcw)\macmaxarcw)^{\macicon'
  \macnumplaces^{4}}$.
\end{lemma}
\begin{proof}
  Let $\macplacesu, \macubs, \macubst \subseteq \macplaces$ be subsets
  of places such that $\macidplaces \subseteq \macplacesu$,
  $|\macplacesu \setminus \macidplaces| = \macplaceidx+1$ and
  $\macubs$ is non-empty. Let $\macmark: \macplaces \to \macint$ be
  some marking. Suppose there is a $\macubst$-neglecting weakly
  $\macmark, \macplacesu, \omega$-enabled $\macubs$-pumping sequence
  $\macfso$ such that every independent place $\macplaceo \in
  \macidplaces \setminus \macubst$ has at most
  $\macmaxinit' + \macidnidx \macmaxarcw$ tokens in any intermediate
  marking belonging to the caring zone of $\macplaceo$. We will prove
  that there is a $\macubst$-neglecting weakly $\macmark, \macplacesu,
  \omega$-enabled $\macubs$-pumping sequence of length at most $10
  \macnumplaces \macvcnum' (2 \macmaxarcw \macpumseqlen( \macplaceidx,
  \macidnidx+1))^{\macicon' \macvcnum'^{3}} ( (\macmaxinit+ \macidnidx
  \macmaxarcw)\macmaxarcw)^{\macicon' \macnumplaces^{4}}$.

  Case 1: The sequence $\macfso$ is a $\macubst$-neglecting weakly
  $\macmark, \macplacesu, \macmaxarcw \macpumseqlen( \macplaceidx,
  \macidnidx+1)$-enabled $\macubs$-pumping sequence. The required
  result is a consequence of \lemref{lem:shortPumSeq}, after
  substituting $\macmaxinit' + \macidnidx \macmaxarcw$ for
  $\macmaxinit'$.

  Case 2: The sequence $\macfso$ decomposes into $\macfso =
  \macfso_{1}' \macpump{ \macfso_{1}} \cdots \macfso_{\macnumpart}'
  \macpump{ \macfso_{\macnumpart}}$ such that for some $2\le
  \macpartidx \le \macnumpart$, $\macmark \macstep{ \macfso_{1}'
  \macpump{ \macfso_{1}} \cdots \macpump{ \macfso_{\macpartidx-1}} }{}
  \macmark_{1} \macstep{\macfso_{\macpartidx}'}{} \macmark_{2}$ and
  there is some intermediate marking $\macmark'$ between
  $\macmark_{1}$ and $\macmark_{2}$ and a place $\macplacet \in
  \macplacesu \setminus \macubst$ with $\macmark' (\macplacet) \ge
  \macmaxarcw \macpumseqlen(\macplaceidx, \macidnidx+1)$. Let
  $\macmark'$ be the earliest such intermediate marking occurring
  outside of pumping portions. If there is some $\macpartidx > 1$ such
  that $\{\macplaceo \in \macplaces \mid \maceff{
  \macpump{\macfso_{\macpartidx}}} (\macplaceo) > 0\} \subseteq
  \bigcup_{\macpartidxt = 1}^{\macpartidx-1}\{\macplaceo \in
  \macplaces \mid \maceff{ \macpump{\macfso_{\macpartidxt}}}
  (\macplaceo) > 0\}$, then $\macpump{ \macfso_{\macpartidx}}$ can be
  considered as a non-pumping portion and the resulting sequence will
  still be a $\macubst$-neglecting weakly $\macmark, \macplacesu,
  \omega$-enabled $\macubs$-pumping sequence. Hence, without loss of
  generality, we can assume that $\macnumpart \le \macnumplaces$. Let
  $\macmark_{1} \macstep{ \macfso_{\macpartidx}^{1\prime}}{} \macmark'
  \macstep{ \macfso_{\macpartidx}^{2\prime}{}}{} \macmark_{2}$. Let
  $\macubs_{1} = \bigcup_{\macpartidxt=1}^{\macpartidx-1}\{\macplaceo
  \in \macplaces \mid \maceff{ \macpump{ \macfso_{\macpartidxt}}}
  (\macplaceo) > 0\}$. The sequence $\macfso_{1}' \macpump{
  \macfso_{1}} \cdots \macpump{ \macfso_{\macpartidx-1}}$ is a
  $\macubst$-neglecting weakly $\macmark, \macplacesu, \macmaxarcw
  \macpumseqlen(\macplaceidx, \macidnidx+1)$-enabled
  $\macubs_{1}$-pumping sequence in which every place $\macplaceo
  \in \macplacesu \setminus \macubst$ has at most
  $\macmaxinit' + \macidnidx \macmaxarcw$ tokens in all intermediate
  markings belonging to the caring zone of $\macplaceo$. By
  \lemref{lem:shortPumSeq}, there is a $\macubst$-neglecting weakly
  $\macmark, \macplacesu, \omega$-enabled $\macubs_{1}$-pumping sequence $\macfst_{1}$
  of length at most $8 (\macpartidx - 1) \macvcnum' (2 \macmaxarcw
  \macpumseqlen(\macplaceidx, \macidnidx+1))^{\macicon'\macvcnum'^{3}} (
  (\macmaxinit' + \macidnidx \macmaxarcw) \macmaxarcw)^{\macicon'
  \macnumplaces^{4}}$. We can remove all $(\macplacesu \setminus
  \macubst \setminus \macubs_{1})$-loops from
  $\macfso_{\macpartidx}^{1\prime}$ to obtain
  $\macfst_{\macpartidx}^{1\prime}$  of length at most
  $(\macmaxarcw \macpumseqlen ( \macplaceidx,
  \macidnidx+1))^{\macvcnum'} (\macmaxinit' +
  \macidnidx \macmaxarcw)^{\macnumplaces}$. If $\macmark \macstep{
  \macfst_{1}}{} \macmark_{1}' \macstep{
  \macfst_{\macpartidx}^{1\prime}} {} \macmark'' \macstep{
  \macfso_{\macpartidx}^{2\prime}}{} \macmark_{2}'$, we will have
  $\macmark''(\macplaceo) = \macmark' (\macplaceo)$ for all
  $\macplaceo \in (\macplacesu \setminus \macubst \setminus
  \macubs_{1})$.

  The sequence $\macfso_{\macpartidx}^{2\prime} \macpump{
  \macfso_{\macpartidx}} \cdots \macfso_{\macnumpart}' \macpump{
  \macfso_{\macnumpart}}$ is a $(\macubst \cup
  \macubs_{1})$-neglecting weakly $\macmark', \macplacesu,
  \omega$-enabled $(\macubs \setminus \macubs_{1})$-pumping sequence
  such that every independent place $\macplaceo \in \macidplaces
  \setminus (\macubst \cup \macubs_{1})$ has at most
  $\macmaxinit' + \macidnidx \macmaxarcw$ tokens in all intermediate
  markings belonging to the caring zone of $\macplaceo$. By
  definition, there is a $(\macubst \cup \macubs_{1})$-neglecting
  weakly $\macmark', \macplacesu \setminus \{\macplacet\},
  \omega$-enabled $(\macubs \setminus \macubs_{1})$-pumping sequence
  $\macfst_{2}$ of length at most $\macpumseqlen( \macplaceidx,
  \macidnidx)$. If $\macplacet \in \macubs_{1}$, then $\macfst_{2}$ is
  also a $(\macubst \cup \macubs_{1})$-neglecting weakly
  $\macmark', \macplacesu, \omega$-enabled $(\macubs \setminus
  \macubs_{1})$-pumping sequence.  Otherwise,
  $\macmark''(\macplacet) = \macmark' (\macplacet) \ge
  \macmaxarcw \macpumseqlen( \macplaceidx, \macidnidx)$ and
  $\macfst_{2}$ can decrease at most $\macmaxarcw \macpumseqlen(
  \macplaceidx, \macidnidx)$ tokens from $\macplacet$, so again
  $\macfst_{2}$ is a $(\macubst \cup \macubs_{1})$-neglecting weakly
  $\macmark', \macplacesu, \omega$-enabled $(\macubs \setminus
  \macubs_{1})$-pumping sequence. In either case,
  \lemref{lem:weakPumpSeqCombine} implies that $\macfst_{1}
  \macfst_{\macpartidx}^{1\prime} \macfst_{2}$ is a
  $\macubst$-neglecting weakly $\macmark, \macplacesu, \omega$-enabled
  $\macubs$-pumping sequence. Its length is at most $8 \macnumpart
  \macvcnum'  (2 \macmaxarcw \macpumseqlen( \macplaceidx,
  \macidnidx+1))^{\macicon' \macvcnum'^{3}} ( (\macmaxinit' + \macidnidx
  \macmaxarcw)\macmaxarcw)^{\macicon' \macnumplaces^{4}} +
  (\macmaxarcw \macpumseqlen(\macplaceidx, \macidnidx+1))^{\macvcnum'}
  (\macmaxinit' + \macidnidx \macmaxarcw)^{\macnumplaces} +
  \macpumseqlen(\macplaceidx, \macidnidx)$.

  Case 3: The sequence $\macfso$ decomposes into $\macfso =
  \macfso_{1}' \macpump{ \macfso_{1}} \cdots \macfso_{\macnumpart}'
  \macpump{ \macfso_{\macnumpart}}$ such that for some intermediate
  marking $\macmark'$ occurring while firing $\macfso_{1}'$ from
  $\macmark$, there is some place $\macplacet \in \macplacesu
  \setminus \macubst$ such that $\macmark' (\macplacet) \ge
  \macmaxarcw \macpumseqlen( \macplaceidx, \macidnidx)$. Let
  $\macmark'$ be the first such intermediate marking. Let $\macmark
  \macstep{ \macfso_{1}^{1\prime} }{} \macmark'
  \macstep{\macfso_{1}^{2\prime}}{} \macmark_{1}$. Remove all
  $\macplacesu\setminus \macubst$-loops from $\macfso_{1}^{1\prime}$
  to get $\macfst_{1}^{1\prime}$ of length at most $(\macmaxarcw
  \macpumseqlen( \macplaceidx, \macidnidx+1))^{\macvcnum'} (
  \macmaxinit' + \macidnidx \macmaxarcw)^{\macnumplaces}$. In
  addition, $\macmark \macstep{ \macfst_{1}^{1\prime}}{} \macmark''$
  such that $\macmark'' (\macplaceo) = \macmark' (\macplaceo)$ for all
  $\macplaceo \in \macplacesu \setminus \macubst$. The sequence
  $\macfso_{1}^{2\prime} \macpump{ \macfso_{1}} \cdots
  \macpump{ \macfso_{\macnumpart}}$ is a $\macubst$-neglecting weakly
  $\macmark', \macplacesu \setminus \{\macplacet\}, \omega$-enabled
  $\macubs$-pumping sequence such that every independent place
  $\macplaceo \in \macidplaces \setminus \macubst$ has
  at most $\macmaxinit' + \macidnidx \macmaxarcw$ tokens in any
  intermediate marking belonging to the caring zone of $\macplaceo$.
  By definition, there is a $\macubst$-neglecting weakly $\macmark',
  \macplacesu \setminus \{\macplacet\}, \omega$-enabled
  $\macubs$-pumping sequence $\macfst$ of length at most
  $\macpumseqlen(\macplaceidx, \macidnidx)$. Since $\macfst$ can
  decrease at most $\macmaxarcw \macpumseqlen(\macplaceidx,
  \macidnidx)$ tokens from $\macplacet$ and $\macmark' (\macplacet) =
  \macmark'' (\macplacet) \ge \macmaxarcw \macpumseqlen( \macplaceidx,
  \macidnidx)$, $\macfst$ is also a $\macubst$-neglecting weakly $\macmark',
  \macplacesu, \omega$-enabled $\macubs$-pumping sequence.
  Hence, $\macfso_{1}^{1\prime} \macfst$ is a $\macubst$-neglecting
  weakly $\macmark, \macplacesu, \omega$-enabled
  $\macubs$-pumping sequence.

  Case 4: The sequence $\macfso$ decomposes into $\macfso =
  \macfso_{1}' \macpump{ \macfso_{1}} \cdots \macfso_{\macnumpart}'
  \macpump{ \macfso_{\macnumpart}}$ such that for some $1\le
  \macpartidx \le \macnumpart$, $\macmark \macstep{ \macfso_{1}'
  \macpump{ \macfso_{1}} \cdots \macfso_{\macpartidx}' }{}
  \macmark_{1} \macstep{ \macpump{\macfso_{\macpartidx}}}{}
  \macmark_{2}$ and there is some intermediate marking $\macmark'$
  between $\macmark_{1}$ and $\macmark_{2}$ and a place $\macplacet
  \in \macplacesu \setminus \macubst$ with $\macmark' (\macplacet) \ge
  \macmaxarcw \macpumseqlen(\macplaceidx, \macidnidx+1)$. For every
  independent place $\macplaceo \in \macidplaces \setminus \macubst$,
  if $\maceff{ \macpump{\macfso_{\macpartidx}}} (\macplaceo) >
  \macmaxarcw$, transfer to $\macplaceo_{\macplvar}$ the last
  transition in $\macpump{ \macfso_{\macpartidx}}$ that adds tokens to
  $\macplaceo$, where $\macplvar$ is the variety of $\macplaceo$.
  Repeat this until for every $\macplaceo \in \macidplaces \setminus
  \macubst$ with $\maceff{ \macpump{ \macfso_{\macpartidx}}}
  (\macplaceo) > 0$, no more than $\macmaxarcw$ and no less than $1$
  tokens are added by the new pumping portion after the transfers. By
  \lemref{lem:weakPupmSeqRep}, $\macfso_{1}' \macpump{ \macfso_{1}}
  \cdots \macfso_{\macpartidx}' \macfso_{\macpartidx} \macpump{
  \macfso_{\macpartidx}} \cdots \macpump{ \macfso_{\macnumpart}}$ is a
  $\macubst$-neglecting weakly $\macmark, \macplacesu, \omega$-enabled
  $\macubs$-pumping sequence such that every independent place
  $\macplaceo \in \macidplaces \setminus \macubst$ has
  at most $\macmaxinit' + (\macidnidx+1) \macmaxarcw$ tokens in any
  intermediate marking belonging to the caring zone of $\macplaceo$.
  Now, we are back to case 2 or case 3 with $(\macidnidx + 1)$
  replacing $\macidnidx$. \qed
\end{proof}

As earlier, we will denote $\macicon' \macvcnum'^{3}$ by $\maciconh$.
\begin{lemma}
  \label{lem:weakPumSeqLenUpBound}
  $\macpumseqlen(\macplaceidx,\macidnidx)\le
  (10 \macnumplaces \macvcnum')^{(1+\maciconh)^{\macplaceidx}}
  (2\macmaxarcw)^{\macpoly_{1}(\maciconh^{\macplaceidx})}
  (\macmaxinit'+(\macidnidx+\macplaceidx)\macmaxarcw)^{\macpoly_{2}(\maciconh^{\macplaceidx})}$
  where $\macpoly_{1}(\maciconh^{\macplaceidx})$ and
  $\macpoly_{2}(\maciconh^{\macplaceidx})$ are polynomials in
  $\maciconh^{\macplaceidx},\macicon',\macvcnum'$ and $\macnumplaces$.
\end{lemma}
\begin{proof}
  By induction on $\macplaceidx$. $\macpumseqlen(0,\macidnidx)\le
  8 \macnumplaces \macvcnum'(2 (\macmaxinit' +\macidnidx
  \macmaxarcw)\macmaxarcw)^{\macicon' \macnumplaces^{4}}$. We will
  choose $poly_{1}$ and $poly_{2}$ such that  $8 \macnumplaces \macvcnum'(2 (\macmaxinit' +\macidnidx
  \macmaxarcw)\macmaxarcw)^{\macicon' \macnumplaces^{4}} \le
  10 \macnumplaces
  \macvcnum'(2 \macmaxarcw)^{poly_{1}(1)} (\macmaxinit'+\macidnidx\macmaxarcw)^{poly_{2}(1)}$.
  \begin{align*}
    \macscovlen(\macplaceidx+1,\macidnidx) &\le 10 \macnumplaces \macvcnum'
    (2\macmaxarcw\macpumseqlen(\macplaceidx,\macidnidx+1))^{\maciconh}
    ( (\macmaxinit'+\macidnidx\macmaxarcw)\macmaxarcw
    )^{\macicon'\macnumplaces^{4}}\\
    &\le 10 \macnumplaces \macvcnum'\left[2\macmaxarcw
    (10 \macnumplaces \macvcnum')^{(1+\maciconh)^{\macplaceidx}}
    (2\macmaxarcw)^{\macpoly_{1}(\maciconh^{\macplaceidx})}
    (\macmaxinit'+(\macidnidx+1+\macplaceidx)\macmaxarcw)^{\macpoly_{2}(\maciconh^{\macplaceidx})}\right]^{\maciconh}\\
    &\quad( (\macmaxinit'+\macidnidx\macmaxarcw)\macmaxarcw
    )^{\macicon'\macnumplaces^{4}}\\
    &\le (10 \macnumplaces \macvcnum')^{1+\maciconh(1+\maciconh)^{\macplaceidx}}
    (2\macmaxarcw)^{(1+\macpoly_{1}(\maciconh^{\macplaceidx}))\maciconh+\macicon'\macnumplaces^{4}}
    (\macmaxinit'+(\macidnidx+\macplaceidx+1)\macmaxarcw)^{\macpoly_{2}(\maciconh^{\macplaceidx})\maciconh
    +\macicon'\macnumplaces^{4}}
  \end{align*}
  It is now enough to choose $\macpoly_{1}$ and $\macpoly_{2}$ such
  that $\macpoly_{1} (\maciconh^{0}) \ge \macicon'
  \macnumplaces^{4}$, $\macpoly_{1}(\maciconh^{\macplaceidx+1})\ge
  (1+\macpoly_{1}(\maciconh^{\macplaceidx}))\maciconh+\macicon'\macnumplaces^{4}$,
  $\macpoly_{2} (\maciconh^{0})\ge \macicon' \macnumplaces^{4}$ and
  $\macpoly_{2}(\maciconh^{\macplaceidx+1})\ge
  \macpoly_{2}(\maciconh^{\macplaceidx})\maciconh +
  \macicon'\macnumplaces^{4}$. These conditions are met by
  $\macpoly_{1}(\maciconh^{\macplaceidx})=\maciconh^{\macplaceidx}
  \macicon' \macnumplaces^{4} +
  (\maciconh+\macicon'\macnumplaces^{4})(\maciconh^{\macplaceidx}-1)$
  and $\macpoly_{2}(\maciconh^{\macplaceidx})=\maciconh^{\macplaceidx}
  \macicon' \macnumplaces^{4} + \macicon'\macnumplaces^{4}(
  \maciconh^{\macplaceidx}-1)$, assuming $\maciconh\ge 2$.\qed
\end{proof}

For the upper bound obtained in \lemref{lem:weakPumSeqLenUpBound} to
be useful, we should have a pumping sequence in which independent
places have controlled number of tokens in intermediate markings
(i.e., $\macmaxinit'$ and $\macidnidx$ are bounded). The
following lemma establishes this with the help of truncation lemma.
\begin{lemma}
  \label{lem:weakPumpSeqIdPlaceFewTokens}
  Let $\macplacesu, \macubs, \macubst \subseteq \macplaces$ be subsets
  of places such that $\macidplaces \subseteq \macplacesu$ and
  $\macubs$ is non-empty. For some $\macmark: \macplaces \to \macint$,
  suppose $\macfso$ is a $\macubst$-neglecting weakly $\macmark,
  \macplacesu, \omega$-enabled $\macubs$-pumping sequence. Let
  $\macmaxinit$ be the maximum of the range of $\macmark$ and let
  $\macmaxinit' = \macmaxinit + \macmaxarcw + \macmaxarcw^{2} +
  \macmaxarcw^{3}$. There is a $\macubst$-neglecting weakly $\macmark,
  \macplacesu, \omega$-enabled $\macubs$-pumping sequence in which
  every independent place $\macplaceo \in \macidplaces \setminus
  \macubst$ has at most $\macmaxinit'$ tokens in all
  intermediate markings belonging to the caring zone of $\macplaceo$.
\end{lemma}
\begin{proof}
  Suppose $\macfso$ is of the form $\macfso = \macfso_{1}' \macpump{
  \macfso_{1}} \macfso_{2}' \macpump{ \macfso_{2}} \cdots
  \macfso_{\macnumpart}' \macpump{ \macfso_{\macnumpart}}$. Ensure
  that for every independent place $\macplaceo \in \macidplaces
  \setminus \macubst$ and $1 \le \macpartidx \le \macnumpart$, if
  $\maceff{ \macpump{ \macfso_{\macpartidx}}} ( \macplaceo) > 0$, then
  $\maceff{ \macpump{ \macfso_{\macpartidx}}} ( \macplaceo) \ge 2
  \macmaxarcw$. If this is not the case, we can repeat $\macpump{
  \macfso_{\macpartidx}}$ $2 \macmaxarcw$ times.
  
  By \lemref{lem:weakPupmSeqRep}, $\macfso_{1}' \macfso_{1} \macpump{
  \macfso_{1}} \macfso_{1} \macfso_{2}' \macfso_{2} \macpump{
  \macfso_{2}} \macfso_{2} \cdots \macfso_{\macnumpart}'
  \macfso_{\macnumpart} \macpump{ \macfso_{\macnumpart}}$ is also a
  $\macubst$-neglecting weakly $\macmark, \macplacesu, \omega$-enabled
  $\macubs$-pumping sequence. Consider some $1\le \macpartidx \le
  \macnumpart$ and an independent place $\macplaceo \in \macidplaces
  \setminus \macubst$ such that $\maceff{ \macpump{
  \macfso_{\macpartidx}}} ( \macplaceo) = 0$ and $\macpump{
  \macfso_{\macpartidx}}$ occurs within the caring zone of
  $\macplaceo$. Let $\macmark \macstep{ \macfso_{1}'\macfso_{1}
  \macpump{ \macfso_{1}} \macfso_{1} \cdots \macfso_{ \macpartidx-1}'
  }{} \macmark_{1} \macstep{\macfso_{\macpartidx}}{} \macmark_{3}
  \macstep{\macpump{ \macfso_{\macpartidx}}}{} \macmark_{4} \macstep{
  \macfso_{\macpartidx}}{} \macmark_{6}$. Let $\macnumtok_{1} = \min
  \{ \macmark'(\macplaceo) \mid \macmark' \text{ occurs between }
  \macmark_{1} \text{ and } \macmark_{3}\}$ be the minimum number of
  tokens in $\macplaceo$ among all intermediate markings occurring
  between $\macmark_{1}$ and $\macmark_{3}$. Let $\macmark_{2}$ be the
  first intermediate marking between $\macmark_{1}$ and $\macmark_{3}$
  such that $\macmark_{2} (\macplaceo) = \macnumtok_{1}$ (see
  \figref{fig:ShortPumSeqProof}).
  \begin{figure}[!htp]
    \begin{center}
      \includegraphics{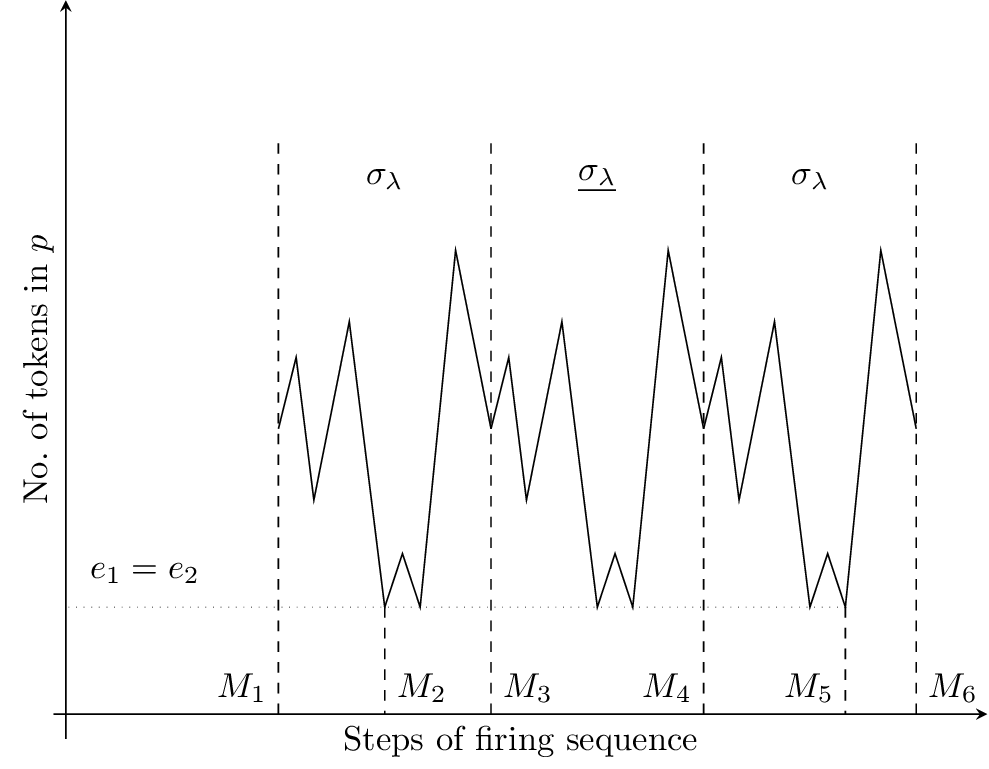}
    \end{center}
    \caption{Illustration for proof of
    \lemref{lem:weakPumpSeqIdPlaceFewTokens}}
    \label{fig:ShortPumSeqProof}
  \end{figure}
  Similarly,
  let $\macnumtok_{2} = \min \{ \macmark'(\macplaceo) \mid \macmark'
  \text{ occurs between } \macmark_{4} \text{ and } \macmark_{6}\}$ be
  the minimum number of tokens in $\macplaceo$ among all intermediate
  markings occurring between $\macmark_{4}$ and $\macmark_{6}$. Let
  $\macmark_{5}$ be the last intermediate marking occurring between
  $\macmark_{4}$ and $\macmark_{6}$ such that $\macmark_{5} (
  \macplaceo) = \macnumtok_{2}$. Note that since $\maceff{ \macpump{
  \macfso_{\macpartidx}}} ( \macplaceo) = \maceff{
  \macfso_{\macpartidx}} ( \macplaceo) = 0$, $\macnumtok_{1} =
  \macnumtok_{2}$. Let $\macmark_{1} \macstep{
  \macfso_{\macpartidx}^{1}}{} \macmark_{2} \macstep{
  \macfso_{\macpartidx}^{2}}{} \macmark_{3} \macstep{ \macpump{
  \macfso_{\macpartidx}}}{} \macmark_{4} \macstep{
  \macfso_{\macpartidx}^{3}}{} \macmark_{5} \macstep{
  \macfso_{\macpartidx}^{4}}{} \macmark_{6}$. Let
  $\macfst_{\macpartidx}$ be the sub-word of
  $\macfso_{\macpartidx}^{2} \macpump{ \macfso_{\macpartidx}}
  \macfso_{\macpartidx}^{3}$ consisting of all the transition
  occurrences having an arc to/from $\macplaceo$. Since $\macmark_{2}
  ( \macplaceo) = \macnumtok_{1} = \macnumtok_{2} = \macmark_{5} (
  \macplaceo)$ is the minimum number of tokens in $\macplaceo$ among
  all intermediate markings occurring between $\macmark_{2}$ and
  $\macmark_{5}$, $\maceff{ \macfst_{\macpartidx}} (\macplaceo) = 0$
  and $\macfst_{\macpartidx}$ is safe for transfer.  Transfer
  $\macfst_{\macpartidx}$ from $\macplaceo$ to
  $\macplaceo_{\macplvar}$, where $\macplvar$ is the variety of
  $\macplaceo$. Perform similar transfers for all $1\le \macpartidx
  \le \macnumpart$ and independent places $\macplaceo \in \macidplaces
  \setminus \macubst$ such that $\maceff{ \macpump{
  \macfso_{\macpartidx}}} ( \macplaceo) = 0$ and $\macpump{
  \macfso_{\macpartidx}}$ occurs within the caring zone of
  $\macplaceo$.

  Consider some $1\le \macpartidx \le
  \macnumpart$ and an independent place $\macplaceo \in \macidplaces
  \setminus \macubst$ such that $\maceff{ \macpump{
  \macfso_{\macpartidx}}} ( \macplaceo) > 0$ and $\macpump{
  \macfso_{\macpartidx}}$ occurs within the caring zone of
  $\macplaceo$. Let $\macmark \macstep{ \macfso_{1}'\macfso_{1}
  \macpump{ \macfso_{1}} \macfso_{1} \cdots \macfso_{ \macpartidx-1}'
  }{} \macmark_{1} \macstep{\macfso_{\macpartidx}}{} \macmark_{3}
  \macstep{\macpump{ \macfso_{\macpartidx}}}{} \macmark_{4}$.  Let
  $\macnumtok_{1} = \min \{ \macmark'(\macplaceo) \mid \macmark'
  \text{ occurs between } \macmark_{1} \text{ and } \macmark_{3}\}$ be
  the minimum number of tokens in $\macplaceo$ among all intermediate
  markings occurring between $\macmark_{1}$ and $\macmark_{3}$. Let
  $\macmark_{2}$ be the first intermediate marking between
  $\macmark_{1}$ and $\macmark_{3}$ such that $\macmark_{2}
  (\macplaceo) = \macnumtok_{1}$. Let $\macmark_{1} \macstep{
  \macfso_{\macpartidx}^{1}}{} \macmark_{2} \macstep{
  \macfso_{\macpartidx}^{2}}{} \macmark_{3} \macstep{ \macpump{
  \macfso_{\macpartidx}}}{} \macmark_{4}$. Let $\macfst_{
  \macpartidx}$ be the sub-word of $\macfso_{\macpartidx}^{2}
  \macpump{ \macfso_{\macpartidx}}$ consisting of all transition
  occurrences having an arc to/from $\macplaceo$. Since
  $\macmark_{2} ( \macplaceo) = \macnumtok_{1}$ is the minimum number
  of tokens in $\macplaceo$ among all intermediate markings between
  $\macmark_{1}$ and $\macmark_{4}$, $\macfst_{\macpartidx}$ is safe
  for transfer. Transfer $\macfst_{\macpartidx}$ to
  $\macplaceo_{\macplvar}$. To ensure that after this transfer, number
  of tokens in $\macplaceo$ is pumped up during the pumping portion
  under consideration, identify the last transition in
  $\macfst_{\macpartidx}$ that adds tokens to $\macplaceo$ and
  transfer it back to $\macplaceo$. Since $\maceff{ \macpump{
  \macfso_{\macpartidx}}} ( \macplaceo) \ge 2 \macmaxarcw$, this last
  back transfer will not violate any property of the pumping sequence.
  Perform this transfer and back transfer for all $1\le \macpartidx \le
  \macnumpart$ and independent places $\macplaceo \in \macidplaces
  \setminus \macubst$ such that $\maceff{ \macpump{
  \macfso_{\macpartidx}}} ( \macplaceo) > 0$ and $\macpump{
  \macfso_{\macpartidx}}$ occurs within the caring zone of
  $\macplaceo$.
  
  Now, we have a $\macubst$-neglecting weakly $\macmark, \macplacesu,
  \omega$-enabled $\macubs$-pumping sequence with the following
  properties:
  \begin{enumerate}
    \item For all $1\le \macpartidx \le \macnumpart$ and independent
      places $\macplaceo \in \macidplaces \setminus \macubst$ such
      that $\maceff{ \macpump{ \macfso_{\macpartidx}}} ( \macplaceo) =
      0$ and $\macpump{ \macfso_{\macpartidx}}$ occurs within the
      caring zone of $\macplaceo$, no transition in $\macpump{
      \macfso_{\macpartidx}}$ has an arc to/from $\macplaceo$.
    \item For all $1\le \macpartidx \le \macnumpart$ and independent
      places $\macplaceo \in \macidplaces \setminus \macubst$ such
      that $\maceff{ \macpump{ \macfso_{\macpartidx}}} ( \macplaceo) >
      0$ and $\macpump{ \macfso_{\macpartidx}}$ occurs within the
      caring zone of $\macplaceo$, there is only one transition in
      $\macpump{\macfso_{\macpartidx}}$ that has an arc to/from
      $\macplaceo$ and this transition adds some tokens to
      $\macplaceo$.
  \end{enumerate}
  Consider an independent place $\macplaceo \in \macidplaces \setminus
  \macubst$ of some variety $\macplvar$. Let $\macmark'$ be the last
  intermediate marking in the caring zone of $\macplaceo$ such that
  $\macmark'(\macplaceo)$ is the minimum number of tokens in
  $\macplaceo$ among all intermediate markings in the caring zone of
  $\macplaceo$.

  Case 1: $\macmark'(\macplaceo) \ge \macmark(\macplaceo)$. In this
  case, the number of tokens in $\macplaceo$ does not come below
  $\macmark ( \macplaceo)$ at all. Let $\macfst_{\macplaceo}$ be the
  sub-word of the pumping sequence consisting of all transitions
  occurrences within the caring zone of $\macplaceo$ that have an arc
  to/from $\macplaceo$, except the last such transition. Transfer
  $\macfst_{\macplaceo}$ to $\macplaceo_{v}$.

  Case 2: $\macmark' ( \macplaceo) < \macmark ( \macplaceo)$. Invoking
  truncation lemma with $\macnumtok = \macmark(\macplaceo) +
  \macmaxarcw$, we identify sub-words between $\macmark$ and
  $\macmark'$ and transfer them to $\macplaceo_{\macplvar}$ so that in
  any intermediate marking within the caring zone of $\macplaceo$,
  $\macplaceo$ has at most $\macmaxinit + \macmaxarcw +
  \macmaxarcw^{2} + \macmaxarcw^{3}$ tokens. Note that none of the
  sub-words transferred will involve any transition in pumping
  portions due to the property we have ensured above.

  Due to the property we have ensured above, if for some place
  $\macplaceo \in \macidplaces \setminus \macubst$, there is some
  $\macpump{ \macfso_{\macpartidxt}}$ occurring within the caring zone
  of $\macplaceo$ with $\maceff{ \macpump{ \macfso_{\macpartidxt}}}
  (\macplaceo) > 0$, it remains so after any of the transfers above.
  For every independent place $\macplaceo \in \macidplaces \setminus
  \macubst$, we identify and transfer sub-words to
  $\macplaceo_{\macplvar}$ based on one of the above two cases.
  Finally, we end up with a $\macubst$-neglecting weakly $\macmark, \macplacesu,
  \omega$-enabled $\macubs$-pumping sequence such that every
  independent place $\macplaceo \in \macidplaces \setminus \macubst$
  has at most $\macmaxinit'$ tokens in all intermediate
  markings belonging to the caring zone of $\macplaceo$. \qed
\end{proof}

We will now combine results of previous lemmas to give a
\macparapsp{} upper bound for model checking $\beta$ formulas.
\begin{theorem}
  \label{thm:modelCheckParapsp}
  With the vertex cover number $\macvcnum$ and maximum arc weight
  $\macmaxarcw$ as parameters, $\beta$ formulas of the logic given in
  the beginning of this section can be model checked in \macparapsp.
\end{theorem}
\begin{proof}
  From \lemref{lem:pumpSeqNecSuf}, model checking $\beta$ formulas is
  equivalent to checking the presence of $\macubs$-pumping sequences
  for some $\macubs$. The choice of $\macubs$ can be done
  non-deterministically in the algorithm. From
  \lemref{lem:WeakEnSeqSuf}, checking the presence of
  $\macubs$-pumping sequences is equivalent to checking the presence
  of $\emptyset$-neglecting weakly $\macmark_{0}, \macplaces,
  \omega$-enabled $\macubs$-pumping sequences. Setting $\macmaxinit' =
  \macmaxinit + \macmaxarcw^{2} + \macmaxarcw^{3}$ in
  \defref{def:pumLenBound}, \lemref{lem:weakPumpSeqIdPlaceFewTokens}
  implies that if there is a $\emptyset$-neglecting weakly
  $\macmark_{0}, \macplaces, \omega$-enabled $\macubs$-pumping
  sequence, there is one of length at most $\macpumseqlen(\macvcnum',
  1)$.

  A non-deterministic Turing machine can test for the presence of a
  weakly enabled pumping sequence by guessing and verifying a sequence
  of length at most $\macpumseqlen(\macvcnum',1)$. By
  \lemref{lem:weakPumSeqLenUpBound}, the memory needed by such a
  Turing machine is $\macOh(\macnumplaces \log|\macmark_{0}|+
  \macnumplaces + \log \macmaxarcw +
  (1+\macicon'\macvcnum'^{3})^{\macvcnum'}\log \macvcnum' \log
  \macnumplaces +
  \macpoly_{1}(\macicon'^{\macvcnum'}\macvcnum'^{3\macvcnum'})\log
  \macmaxarcw +
  \macpoly_{2}(\macicon'^{\macvcnum'}\macvcnum'^{3\macvcnum'})
  \log(\macmaxinit'\macvcnum'\macmaxarcw))$, or $\macOh(\macnumplaces
  \log|\macmark_{0}| + \macnumplaces +
  \macpoly(\macicon'^{3\macvcnum'}
  \macvcnum'^{3\macvcnum'})\log(\macmaxinit'\macvcnum' \macnumplaces
  \macmaxarcw))$ for some polynomial $\macpoly$.  An application of
  Savitch's theorem now gives us the required \macparapsp{} algorithm.
  \qed
\end{proof}
\end{full}

\section{Conclusion}
With the vertex cover number of the underlying graph of a Petri net
and maximum arc weight as parameters, we proved that the coverability
and boundedness problems can be solved in \macparapsp{}. A fragment of
CTL based on these two properties can also be model checked in
\macparapsp{}. Since vertex cover is better studied than the
parameter benefit depth we introduced in \cite{PL09}, the results here
might lead us towards applying other techniques of parameterized
complexity to these problems. Whether coverability and boundedness are
in \macparapsp{} with the size of the smallest feedback vertex set and
maximum arc weight as parameters is an open problem.

\textbf{Acknowledgements.} The author acknowledges Kamal Lodaya and
Saket Saurabh for helpful discussions and feedback on the draft.

\bibliographystyle{plain}
\bibliography{references}

\begin{full}
\newpage
\appendix
\section{Proof of Truncation Lemma}
\begin{proof}[\lemref{lem:truncationLemma}]
  Let $\macmark_{1}'$ be the last intermediate marking before
  $\macmark_{2}$ such that $\macmark_{1}'(\macplaceo_{1})\le
  \macnumtok+\macmaxarcw^{2}$ (see \figref{fig:truncLemmaProof}).
  \begin{figure}[!htp]
    \begin{center}
      \includegraphics{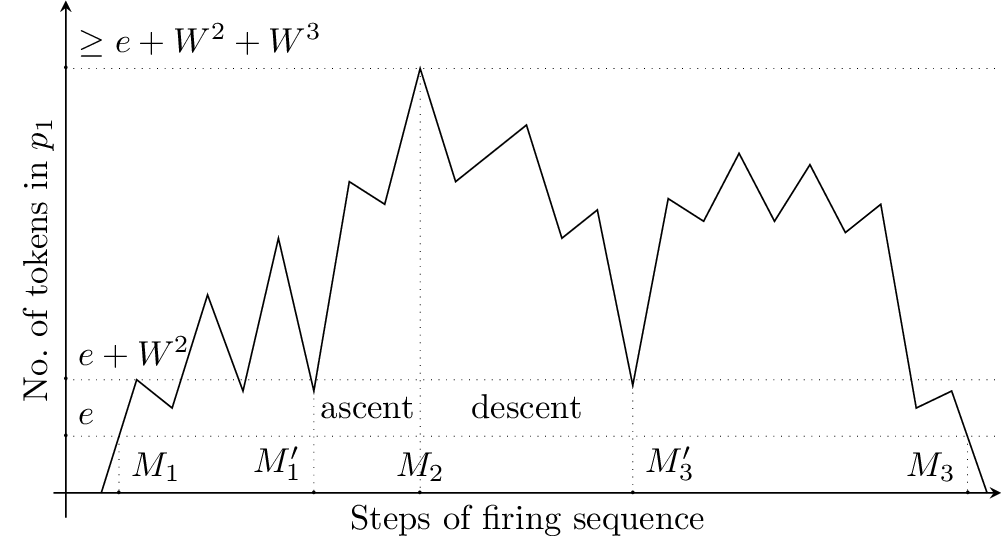}
    \end{center}
    \caption{Illustration for proof of \lemref{lem:truncationLemma}}
    \label{fig:truncLemmaProof}
  \end{figure}
  Let $\macmark_{3}'$ be the first intermediate marking after
  $\macmark_{2}$ such that $\macmark_{3}'(\macplaceo_{1})\le
  \macnumtok+\macmaxarcw^{2}$. We will call the subsequence between
  $\macmark_{1}'$ and $\macmark_{2}$ as \macpopterm{ascent} and the
  subsequence between $\macmark_{2}$ and $\macmark_{3}'$ as
  \macpopterm{descent}. During ascent, the number of tokens in
  $\macplaceo_{1}$ increases by at least $\macmaxarcw^{3}$. Since each
  transition can add at most $\macmaxarcw$ tokens to $\macplaceo_{1}$,
  there are at least $\macmaxarcw^{2}$ transitions adding tokens to
  $\macplaceo_{1}$ during ascent. There must be at least one number
  $1\le \macarcw_{1} \le \macmaxarcw$ such that among these $\macmaxarcw^{2}$
  transitions, there are at least $\macmaxarcw$ transitions that
  add exactly $\macarcw_{1}$ tokens to $\macplaceo_{1}$. Similarly,
  there is a number $1\le \macarcw_{2}\le \macmaxarcw$ such that at
  least $\macmaxarcw$ transitions remove exactly $\macarcw_{2}$ tokens
  from $\macplaceo_{1}$ during descent. The sub-word $\macfirseqo'$
  we need consists of $\macarcw_{2}$ ``adding'' transitions from
  ascent and $\macarcw_{1}$ ``removing'' transitions from descent.
  The total effect of $\macfirseqo'$ on $\macplaceo_{1}$ is $0$ and it
  is safe for transfer from $\macplaceo_{1}$ to $\macplaceo_{2}$ by
  construction. Since the first part of $\macfirseqo'$ removes
  $\macarcw_{1}\macarcw_{2}>0$ tokens from $\macplaceo_{1}$, the
  number of tokens $\macmark_{2}(\macplaceo_{1})$ after transferring
  $\macfirseqo'$ to $\macplaceo_{2}$ is strictly less than the number
  of tokens before the transfer. Before transfer, every intermediate
  marking between $\macmark_{1}'$ and $\macmark_{3}'$ had at least
  $\macnumtok+\macmaxarcw^{2}$ tokens.  Since the transfer of
  $\macfirseqo'$ causes $\macarcw_{1}\macarcw_{2}\le \macmaxarcw^{2}$
  fewer tokens, all intermediate markings between $\macmark_{1}'$ and
  $\macmark_{3}'$ will have at least $\macnumtok\ge 0$ tokens in
  $\macplaceo_{1}$ after transfer. Intermediate markings before
  $\macmark_{1}'$ and after $\macmark_{3}'$ do not change.\qed
\end{proof}
\end{full}

\end{document}